\documentclass{scrartcl}
\usepackage[left=2cm, right=2cm, top=2cm, bottom=3cm]{geometry}
\usepackage[utf8]{inputenc}
\usepackage{graphicx}
\usepackage{siunitx,physics}
\usepackage{hyperref}
\usepackage{cleveref}
\usepackage[affil-it]{authblk}

\usepackage[backend=biber,style=phys,maxnames=10]{biblatex}
\usepackage{bm, booktabs}
\newcommand{\targetsens}{$10^{-29}~ e \cdot \textnormal{cm}$}
\title{The storage ring proton EDM experiment}
\date{April 20, 2022}
\author[7]{Jim Alexander}
\author[36]{Vassilis Anastassopoulos}
\author[28]{Rick Baartman}
\author[39,22]{Stefan Bae\ss{}ler}
\author[19]{Franco Bedeschi}
\author[17]{Martin Berz}
\author[4]{Michael Blaskiewicz}
\author[33]{Themis Bowcock}
\author[4]{Kevin Brown}
\author[9,31]{Dmitry Budker}
\author[33]{Sergey Burdin}
\author[8]{Brendan C. Casey}
\author[34]{Gianluigi Casse}
\author[38]{Giovanni Cantatore}
\author[34]{Timothy Chupp}
\author[4]{Hooman Davoudiasl}
\author[4]{Dmitri Denisov}
\author[4]{Milind V. Diwan}
\author[20]{George Fanourakis}
\author[30,36]{Antonios Gardikiotis}
\author[18]{Claudio Gatti}
\author[33]{James Gooding}
\author[32]{Renee Fatemi}
\author[4]{Wolfram Fischer}
\author[26]{Peter Graham}
\author[23]{Frederick Gray}
\author[6]{Selcuk Haciomeroglu}
\author[7]{Georg H. Hoffstaetter}
\author[4]{Haixin Huang}
\author[19]{Marco Incagli}
\author[16]{Hoyong Jeong}
\author[13]{David Kaplan}
\author[37]{Marin Karuza}
\author[29]{David Kawall}
\author[6]{On Kim}
\author[5]{Ivan Koop}
\author[14,8]{Valeri Lebedev}
\author[27]{Jonathan Lee}
\author[6]{Soohyung Lee}
\author[25,19]{Alberto Lusiani}
\author[4]{William J. Marciano}
\author[36]{Marios Maroudas}
\author[6]{Andrei Matlashov}
\author[4]{Francois Meot}
\author[3]{James P. Miller}
\author[4]{William M. Morse}
\author[3,8]{James Mott}
\author[15,6]{Zhanibek Omarov}
\author[11]{Cenap Ozben}
\author[6]{SeongTae Park}
\author[35]{Giovanni Maria Piacentino}
\author[4]{Boris Podobedov}
\author[12]{Matthew Poelker}
\author[39]{Dinko Pocanic}
\author[33]{Joe Price}
\author[4]{Deepak Raparia}
\author[13]{Surjeet Rajendran}
\author[4]{Sergio Rescia}
\author[3]{B. Lee Roberts}
\author[6,15]{Yannis K. Semertzidis \thanks{Corresponding author, semertzidisy@gmail.com}}
\author[14]{Alexander Silenko}
\author[4]{Amarjit Soni}
\author[10]{Edward Stephenson}
\author[12]{Riad Suleiman}
\author[21]{Michael Syphers}
\author[24]{Pia Thoerngren}
\author[4]{Volodya Tishchenko}
\author[4]{Nicholaos Tsoupas}
\author[1]{Spyros Tzamarias}
\author[18]{Alessandro Variola}
\author[19]{Graziano Venanzoni}
\author[33]{Eva Vilella}
\author[33]{Joost Vossebeld}
\author[2]{Peter Winter}
\author[16]{Eunil Won}
\author[4]{Anatoli Zelenski}
\author[36]{Konstantin Zioutas}

\affil[1]{%
  Aristotle University of Thessaloniki, Thessaloniki, Greece}
\affil[2]{%
  Argonne National Laboratory, Lemont, Illinois, USA}
\affil[3]{%
  Boston University, Boston, Massachusetts, USA}
\affil[4]{%
  Brookhaven National Laboratory, Upton, New York, USA}
\affil[5]{%
  Budker Institute of Nuclear Physics, Novosibirsk, Russia}
\affil[6]{%
  Center for Axion and Precision Physics Research, Institute for Basic Science, Daejeon, 	Korea}
\affil[7]{%
  Cornell University, Ithaca, New York, USA}
\affil[8]{%
  Fermi National Accelerator Laboratory, Batavia, Illinois, USA}
\affil[9]{%
  Helmholtz-Institute Mainz, Johannes Gutenberg University, Mainz, Germany}
\affil[10]{%
  Indiana University, Bloomington, Indiana, USA}
\affil[11]{%
  Istanbul Technical University, Istanbul, Turkey}
\affil[12]{%
  JLAB, Newport News, Virginia, USA}
\affil[13]{%
  Johns Hopkins University, Baltimore, Maryland, USA}
\affil[14]{%
  Joint Institute for Nuclear Research, Dubna, Russia}
\affil[15]{%
  Physics Dept., KAIST, Daejeon, Korea}
\affil[16]{%
  Physics Dept., Korea University, Seoul, Korea}
\affil[17]{%
  Michigan State University, East Lansing, Michigan, USA}
\affil[18]{%
  National Institute for Nuclear Physics (INFN-Frascati), Rome, Italy}
\affil[19]{%
  National Institute for Nuclear Physics (INFN-Pisa), Pisa, Italy}
\affil[20]{%
  NCSR Demokritos Institute of Nuclear and Particle Physics, Athens, Greece}
\affil[21]{%
  Northern Illinois University, DeKalb, Illinois, USA}
\affil[22]{%
  Oak Ridge National Laboratory, Oak Ridge, TN, USA}
\affil[23]{%
  Regis University, Denver, Colorado, USA}
\affil[24]{%
  Royal Institute of Technology, Division of Nuclear Physics, Stockholm, Sweden}
\affil[25]{%
  Scuola Normale Superiore di Pisa, Pisa, Italy}
\affil[26]{%
  Stanford University, Stanford, California, USA}
\affil[27]{%
  Stony Brook University, Stony Brook, New York, USA}
\affil[28]{%
  TRIUMF, Vancouver, British Columbia, Canada}
\affil[29]{%
  UMass Amherst, Amherst, Massachusetts, USA}
\affil[30]{%
  Universität Hamburg, Hamburg, Germany}
\affil[31]{%
  University of California at Berkeley, Berkeley, California, USA}
\affil[32]{%
  University of Kentucky, Lexington, Kentucky, USA}
\affil[33]{%
  University of Liverpool, Liverpool, UK}
\affil[34]{%
  University of Michigan, Ann Arbor, Michigan, USA}
\affil[35]{%
  University of Molise, Campobasso, Italy}
\affil[36]{%
  University of Patras, Dept. of Physics, Patras-Rio, Greece}
\affil[37]{%
  University of Rijeka, Rijeka, Croatia}
\affil[38]{%
  University of Trieste and National Institute for Nuclear Physics (INFN-Trieste), Trieste, Italy}
\affil[39]{%
  University of Virginia, Charlottesville, Virginia, USA}

\begin{document}
\pagenumbering{roman}

\maketitle
\begin{abstract}
We describe a proposal to search for an intrinsic electric dipole moment (EDM) of the proton with a sensitivity of \targetsens, based on the vertical rotation of the polarization of a stored proton beam. The New Physics reach is of order $\SI{1e3}{TeV}$ mass scale. Observation of the proton EDM provides the best probe of CP-violation in the Higgs sector, at a level of sensitivity that may be inaccessible to electron-EDM experiments. The improvement in the sensitivity to $\theta_{QCD}$, a parameter crucial in axion and axion dark matter physics, is about three orders of magnitude.

\end{abstract}

\pagebreak
\section*{One Page Summary of the Storage Ring Proton EDM Experiment}

\begin{center}
    \includegraphics[width=5cm]{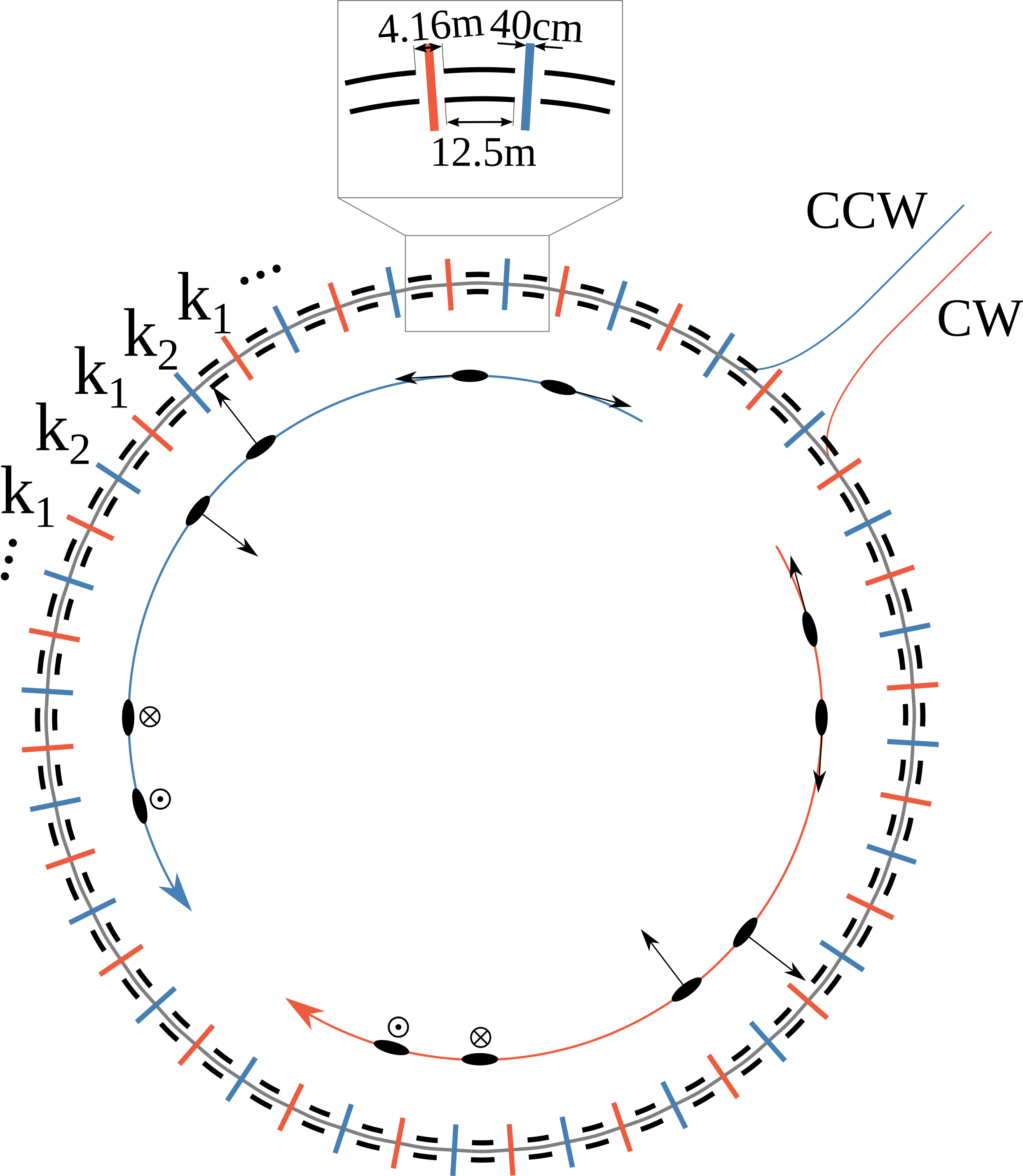}
\end{center}
\pagenumbering{arabic}
\begin{itemize}
\item Proton EDM sensitivity \targetsens. 
\item Improves the sensitivity to QCD CP-violation ($\theta_{\textnormal{QCD}}$) by three orders of magnitude, currently set by the neutron EDM experimental limits.
\item New Physics reach is of order $\SI{1e3}{TeV}$ mass scale~\cite{symmetric}.
\item Probes CP-violation in the Higgs sector with best sensitivity~\cite{edmtheory}.
\item Highly symmetric, magic momentum storage ring lattice in order to control systematics.
    \begin{itemize}
        \item Proton magic momentum =$\SI{0.7}{GeV/c}$.
        \item Proton polarimetry peak sensitivity at the magic momentum.
        \item Optimal electric bending and magnetic focusing.
        \item $\num{2e10}$ polarized protons per fill. One fill every twenty minutes.
        \item Simultaneously stores clockwise (CW) and counterclockwise (CCW) bunches.
        \item Simultaneously stores longitudinally polarized bunches with positive and negative helicities as well as radially polarized bunches.
        \item 24-fold symmetric storage ring lattice.
        \item Changes sign of the focusing/defocusing quadrupoles within 0.1\% of ideal current setting per flip.
        \item Keeps the vertical spin precession rate low when the beam planarity is within $\SI{0.1}{mm}$ over the whole circumference and the maximum split between the counter-rotating (CR) beams is $<\SI{0.01}{mm}$.
        \item Closed orbit automatically compensates spin precession from radial magnetic fields.
        \item Circumference = $\SI{800}{m}$ with $E=\SI{4.4}{MV/m}$, a conservative electric field strength. 
    \end{itemize}
\item 3 -- 5 years of construction and 2 -- 3 years (for statistics collection) to first physics publication. 
\item Sensitive to  dark matter, vector dark matter/dark energy (DM/DE) models~\cite{graham_paper,kim_new_2021}.
  DM/DE signal proportional to $\beta=v/c$. Magic momentum pEDM ring $\beta=0.6$.
\item pEDM is highly complementary to atomic and molecular (AMO) EDM experiments~\cite{edmtheory2}.
  AMO: many different effects, “sole source analysis”, unknown cancellations~\cite{RevModPhys.91.015001}.
\item After proton EDM, can add magnetic bending for deuteron/$^3$He EDM measurements.
  Deuteron and $^3$He EDM measurements complementary physics to proton EDM.

\end{itemize}
\pagebreak
\section*{History}
 The proposed method has its origins in the measurements of the anomalous magnetic moment of the muon in the 1950-70s at CERN. The CERN I experiment~\cite{cern_report} was limited by statistics. The sensitivity breakthrough was to go to a magnetic storage ring. The CERN II result was then limited by the systematics of knowing the magnetic field seen by the muons in the quadrupole magnet. The CERN III experiment~\cite{cern_report,cern3} used an ingenious method to overcome this. It was realized that an electric field at the so-called ``magic'' momentum does not influence the particle $(g-2)$ precession. Rather, the electric field precesses the momentum and the spin at exactly the same rate, so the difference is zero. The fact that all electric fields have this feature, opened up  the possibility of using electric quadrupoles in the ring to focus the beam, while the magnetic field is kept uniform. 
 
 The precession rate of the longitudinal component of the spin in a storage ring with electric and magnetic fields is given by:
\begin{equation}
    \dv{\bm{\beta} \cdot \bm{s}}{t} = -\frac{e}{m} \bm{s}_\perp  \cdot \qty[\qty(\frac{g-2}{2}) \hat{\beta} \times \bm{B} + \qty(\frac{g\beta}{2} - \frac{1}{\beta})\frac{\bm{E}}{c}].
    \label{eq:omega}
\end{equation}

The CERN III experiment used a bending magnetic field with electric quadrupoles for focusing at the ``magic'' momentum, given by $\beta^2=2/g$; see \Cref{eq:omega} electric field term. The CERN III experiment and the BNL version of it, E821~\cite{bennett_final_2006}, were limited by statistics, not systematics. The recent announcement of the $(g-2)$ experimental results~\cite{fnal1} from Fermilab at $\SI{460}{ppb}$ has confirmed the BNL results, with similar statistical and smaller systematic errors. We believe that the FNAL E989 final results, at about $\SI{140}{ppb}$, will have equal statistical and systematic errors. The storage ring/magic momentum breakthrough gained a factor of \num{2e3} in systematic error. 
BNL E821 set a ``parasitic'' limit on the EDM of the muon: $d_{\mu} < \num{1.9e-19}~ e \cdot \textnormal{cm}$~\cite{bnl_edm}. For FNAL E989, we expect this result to improve by up to two orders of magnitude. The statistical and systematic errors on the muon EDM will then be roughly equal. The dominant systematic error effect is due to radial magnetic fields. 

For the pEDM experiment, we plan to use a storage ring at the proton magic momentum with electric bending and magnetic focusing, which gives a negligible radial magnetic field systematic effect --- see below --- while the dominant (main) systematic errors drop out with simultaneous clockwise and counterclockwise storage.
For both BNL E821 and FNAL E989, new systematic effects were discovered that were not in the original proposals. Several ways were applied to mitigate these small effects so they are not the limiting factors. For the pEDM experiment we can get  $10^{11}$ polarized protons per fill from the BNL LINAC/Booster system, and we use symmetries to handle the systematics down to the level of sensitivity. We expect that at that level we perhaps will also discover new small systematics effects, as in the $(g-2)$ experiments. 

\begin{figure}[!h]
\centering
\includegraphics[width=0.85\textwidth]{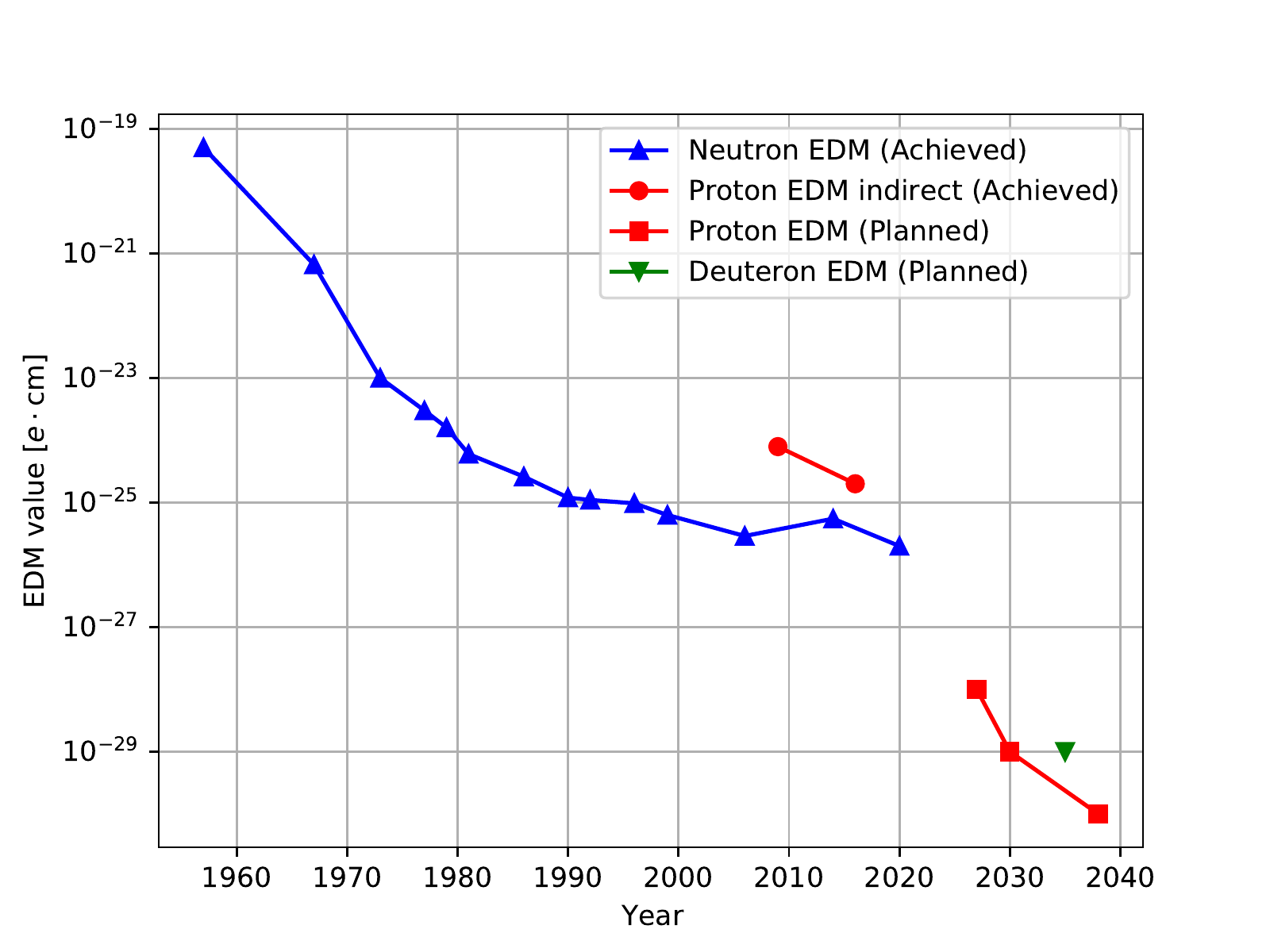}
\caption{The neutron and proton (indirect) EDM limits by publication year are shown here. The storage ring EDM projected sensitivities for the proton and deuteron nuclei are also shown as a function of year. The $^3$He nucleus storage ring EDM sensitivity is projected to be similar to that of the deuteron.}\label{fig:PartEDM}
\end{figure}

Current searches for the EDM of fundamental particles have a large range  of experimental sensitivity, as well as New Physics probing strength.
 Some of the strongest probes of New Physics come from the experimental limits of the electron (inferred indirectly from the atomic ThO EDM limit), the neutron and $^{199}{\textnormal{Hg}}$ EDM limits~\cite{RevModPhys.91.015001,neutronEDM2020,PhysRevLett.97.131801,electron_edm1,electron_edm2,hg_edm,PhysRevLett.119.119901}. Their New Physics reach is the same within one order of magnitude of each other, while the proton EDM at \targetsens ~will bring a more than three orders of magnitude improved sensitivity over the current neutron limit. The current (indirect) experimental limit of the proton at $10^{-25}~ e \cdot \textnormal{cm}$ is derived from the $^{199}{\textnormal{Hg}}$ atomic EDM limit.
In the last three decades there has been a large effort to develop a stronger ultra-cold-neutron (UCN) source, e.g., see~\cite{RevModPhys.91.015001,kuchler_2019,golub75,golub77,shin:2018emy}, to enhance the probability for a higher sensitivity neutron EDM experiment beyond the few $10^{-28} \, e \cdot {\textnormal{cm}}$, the currently best experimental target for the neutron.
  \Cref{fig:PartEDM}  shows the experimental limits of the neutron by publication year, the indirect proton EDM limits from the $^{199}{\textnormal{Hg}}$ atomic EDM limit, and the projected sensitivity levels for the proton and deuteron using the storage ring EDM method. The $^{3}{\textnormal{He}}$ sensitivity level is expected to be similar to that of the deuteron.

\section*{The storage ring EDM method}

\begin{figure}[!hb]
    \centering
    \includegraphics[width=0.8\textwidth]{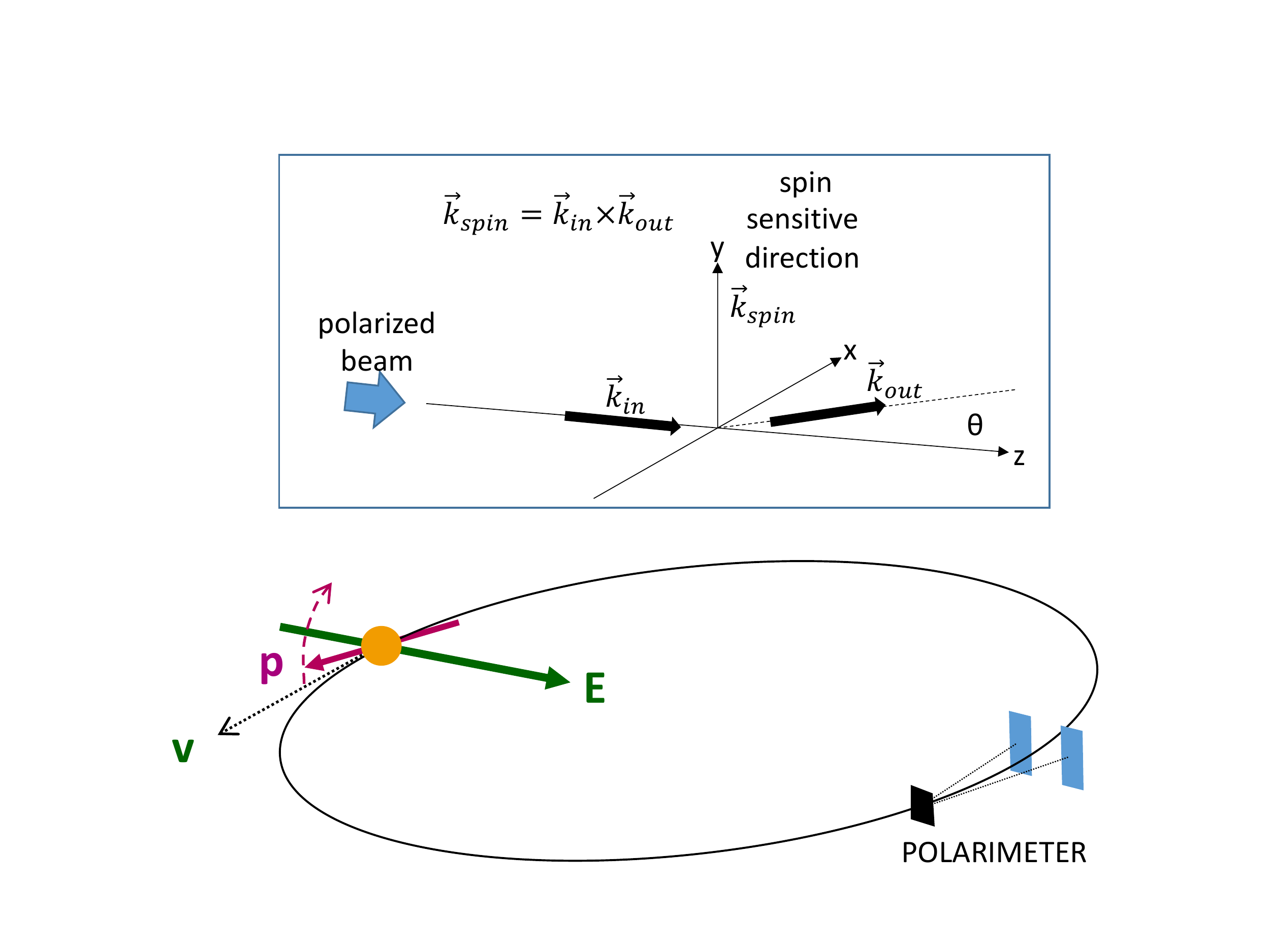}
    \caption{Diagram of the storage ring EDM concept, with the horizontal spin precession locked to the momentum precession rate (``frozen'' spin). The radial electric field acts on the particle EDM for the duration of the storage time.  Positive and negative helicity bunches are stored, as well as bunches with their polarization pointing in the radial direction, for systematic error cancellations. In addition, simultaneous clockwise and counterclockwise storage is used to cancel the main systematic errors. The ring circumference is about 800~m. The top inset shows the cross section geometry that is enhanced in parity-conserving Coulomb and nuclear scattering as the EDM signal increases over time.}
    \label{fig:EJS_ring}
\end{figure}

The concept of the storage ring EDM experiment is illustrated in \Cref{fig:EJS_ring}. There are three starting requirements: (1) The proton beam must be highly polarized in the ring plane. (2) The momentum of the beam must match the magic value of $p=0.7007$~GeV/c, where the ring-plane spin precession is the same as the velocity precession, a condition called ``frozen spin.'' (3) The polarization is initially along the axis of the beam velocity.

The electric field acts along the radial direction toward the center of the ring (E). It is perpendicular to the spin axis (p) and therefore perpendicular to the axis of the EDM. In this situation the spin will precess in the vertical plane as shown in \Cref{fig:EJS_ring}. The appearance of a vertical polarization component with time is the signal for a non-vanishing EDM. This signal is measured at the polarimeter where a sample of the beam is continuously brought to a carbon target. Elastic proton scattering is measured by two downstream detectors (shown in blue)~\cite{hom88,brantjes2012correcting}. The rates depend on the polarization component $p_y$ because it is connected to the {\textit{axial}} vector created from the proton momenta $\vec{k}_{in}\times\vec{k}_{out}$. The sign of $p_y$ flips between left and right as it follows the changing direction of $\vec{k}_{out}$. Thus, the asymmetry in the left-right rates, $(L-R)/(L+R)=p_yA$, is proportional to $p_y$ and hence the magnitude of the EDM. The size of the effect at any given scattering angle also depends on the analyzing power $A$, a property of the scattering process. Having both left and right rates together reduces systematic errors.

A limited number of sensitive storage ring EDM experimental methods have been developed with various degrees of sensitivity and levels of systematic error, see \Cref{tab:lattices}~\cite{farley_new_2004,symmetric}. Here we only address the method based on the hybrid-symmetric ring lattice, which has been studied extensively and shown to perform well, applying
presently available technologies. The other methods, although promising, are outside the scope of this document, requiring additional studies and further technical developments.

The hybrid-symmetric ring method is built on the all-electric ring method, improving it in a number of critical ways that make it practical with present technology. It replaces  electric focusing with alternating gradient magnetic focusing, still allowing simultaneous CW and CCW storage and eliminating the main systematic error source by design. A major improvement in this design is the enhanced ring-lattice symmetry, eliminating the next most-important systematic error source, that of the average vertical beam velocity within the bending sections~\cite{symmetric}.

Symmetries in the hybrid-symmetric ring with \targetsens\ sensitivity:

\begin{enumerate}
    \item CW and CCW beam storage simultaneously.
    \item Longitudinally polarized beams with both helicities.
    \item Radially polarized beams with both polarization directions.
    \item Current flip of the magnetic quadrupoles.
    \item Beam planarity to \SI{0.1}{mm} and beam splitting of the counter-rotating (CR) beams to $<\SI{0.01}{mm}$.
\end{enumerate}

\begin{table*}
    \footnotesize
  \centering
  \caption{Storage ring electric dipole moment experiment options}\label{tab:lattices}  
  \begin{tabular}[t]{p{0.2\linewidth} p{0.12\linewidth} p{0.15\linewidth} p{0.3\linewidth}}
    \toprule
    Fields & Example & EDM signal term & Comments \\ \midrule
    
    Dipole magnetic field $\vb{B}$ (Parasitic).
    & Muon $(g-2)$ experiment.
    & Tilt of the spin precession plane. (Limited statistical sensitivity due to non-zero $(g-2)$ spin precession.) 
    & Eventually limited by geometrical alignment. Requires consecutive CW and CCW injection to eliminate systematic errors. \vspace{1cm}\\
    
    Combination of electric and magnetic fields ($\vb{E}, \vb{B}$) (Combined lattice).
    & Deuteron, $^3\textnormal{He}$, proton.
    & $\dv{\bm{s}}{t} \approx \bm{d}\times\qty(\bm{v}\times\bm{B})$
    & High statistical sensitivity. Requires consecutive CW and CCW injection, with main fields flipping sign to eliminate systematic errors.\vspace{0.5cm}\\
    
    Radial Electric field ($\vb{E}$) and Electric focusing ($\vb{E}$) (All-electric lattice).
    & Proton.
    & $\dv{\bm{s}}{t} = \bm{d}\times\bm{E}$
    & Allows simultaneous CW and CCW storage. Requires demonstration of adequate sensitivity to radial $\bm{B}$-field systematic error source.\vspace{0.5cm}\\
    
    Radial Electric field ($\vb{E}$) and Magnetic focusing ($\vb{B}$)
    (Hybrid, symmetric lattice).
    & Proton.
    & $\dv{\bm{s}}{t} = \bm{d}\times\bm{E}$
    & Allows simultaneous CW and CCW storage. Only lattice to achieve direct cancellation of the main systematic error sources (its own ``co-magnetometer''). \\
    \bottomrule
\end{tabular}
\end{table*}

\section*{Strategy for building a high sensitivity hadronic EDM experiment}
The storage ring EDM method for the proton and deuteron nuclei with frozen spin provides the potential  for a high sensitivity of \targetsens, as explained below in the ``EDM Statistics'' section.  The reason for the high potential sensitivity is the availability of high-intensity, highly-polarized proton and deuteron beams with small phase-space emittance, since they are obtained from polarized ion sources, i.e., a primary source.  Due to the negative value of the deuteron magnetic anomaly, the fields needed for the deuteron case are more complicated than for the proton and the uncertainties are thus larger~\cite{ags_proposal}. The proton EDM ring, using the hybrid-symmetric ring lattice, has been studied extensively~\cite{symmetric}, see also~\cite{anastassopoulos_storage_2016,haciomeroglu_hybrid_2018}, and the conceptual design report (CDR) will be largely based on it. The cost of the experiment is similar to the muon $(g-2)$ cost of about \$\SI{100}{M}.

In preparing for the technical design report (TDR), we will assess the relevant concepts and techniques that have been studied so far~\cite{brantjes2012correcting,morse2013rf,metodiev_fringe_2014,metodiev_analytical_2015,jedi_collaboration_how_2016,hempelmann-phase-2017,SBPM_1,chang2019axionlike,kim_new_2021,graham_paper,bmt3,PRDThomas,PRD2016,PRD2017,PRD2007,OFS,LaszloZimboras,NikVerg}.
We will also:
\begin{enumerate}
    \item Develop prototypes of polarimeters, with an emphasis on minimizing systematic errors and optimizing the statistical power of the method. Test the prototypes for high rates.
  
    \item  Study the optimum material, height, and shape of electric field plates with a field strength of \SI{4.4}{MV/m} for \SI{4}{cm} plate separation, in order 
    to minimize the highest $E$---field value.  Comment: Small surface area $E$---field plates made of aluminum and coated with TiN have been developed and used at J-LAB; rather cheap and robust~\cite{electrode1,electrode2,electrode3}. Need to expand this technology to large-area,  about \SI{20}{cm} high and \SI{2}{m} long.
    
    \item Construct a hydraulic level reference system (HLS) able to keep the ring planarity of the stored beam within \SI{0.1}{mm}. A similar system developed at Fermilab~\cite{shiltsev_space-time_nodate,shiltsev} would be adequate for the needs of the experiment.
    
    \item Test a magnetometer capable of probing the separation of counter-rotating beams by $\SI{10}{\micro m}$. SQUID-based magnetometers have demonstrated \SI{10}{nm \per \sqrt{Hz}} in the lab~\cite{SBPM_1} much better than needed; cheaper technologies are also available.
    
    \item Develop a magnetic quadrupole prototype with emphasis on systematic error minimization when flipping the currents. 
    
    \item Design and construct a combined (hybrid) sextupole system including electric and magnetic fields. 
    
    \item Study the application of trim fields, both electric and magnetic, and develop prototypes of both.
    
    \item Produce a detailed study of the RF-cavity, including a choice of the frequency and tunable range.
    
    \item Construct a straight section equal to 1/48th of the ring and operate all elements together to discover any possible interferences. 
\end{enumerate}

\section*{A Highly Symmetric Lattice}
\begin{figure}
\centering
\includegraphics[width=1.\textwidth]{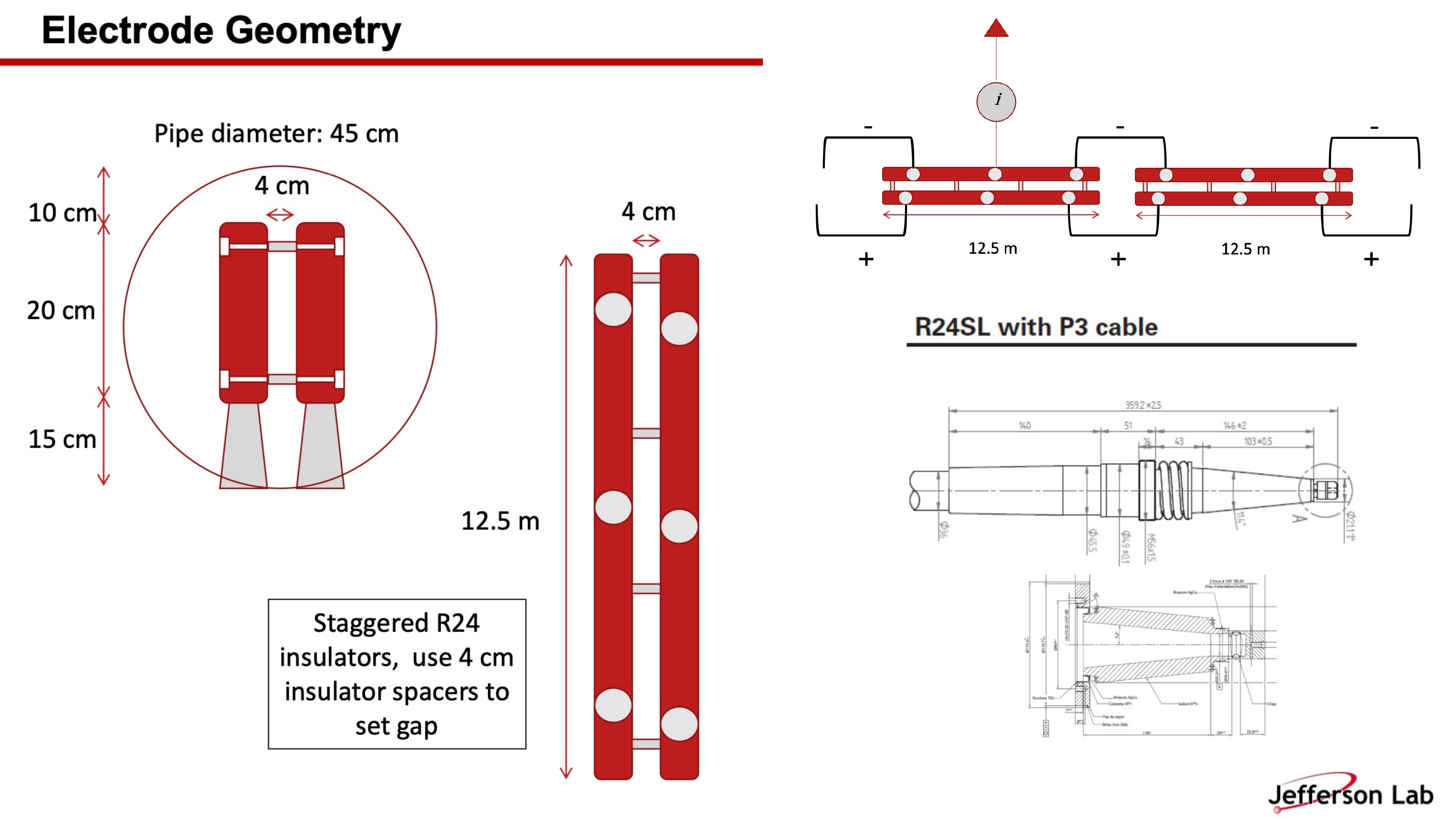}
\caption{The cross-sectional and top views of the electric field plate design under consideration.}\label{fig:Efield_plates}
\end{figure}

A highly symmetric lattice is necessary to limit the EDM and dark matter/dark energy systematics, see~\cite{graham_paper,symmetric}. The 24-fold symmetric ring parameters are given in \Cref{tab:specs}. The ring circumference is \SI{800}{m}, with bending electric field \SI{4.4}{MV/m}. This circumference is that of the BNL AGS tunnel, which would save tunnel construction costs. $E=\SI{4.4}{MV/m}$ is conservative. A pEDM experiment at another location could have up to $E=\SI{5}{MV/m}$ without R\&D progress, see~\cite{electrode1,electrode2,electrode3} and \Cref{fig:Efield_plates}, setting the scale of the required ring circumference.

\begin{table}[tbp]
  \centering
  \caption{Ring and beam parameters for the hybrid-symmetric ring design. The beam planarity refers to the average vertical orbit of the counter-rotating (CR) beams with respect to gravity around the ring.}
  \begin{tabular}[t]{lc}
    \toprule
    Quantity                                                      & Value                        \\
    \midrule
    Bending Radius $R_{0}$                                        & \SI{95.49}{m}                \\
    Number of periods                                             & 24                           \\
    Electrode spacing                                             & \SI{4}{cm}                   \\
    Electrode height                                              & \SI{20}{cm}                  \\
    Deflector shape                                               & cylindrical                  \\
    Radial bending $E$-field                                      & \SI{4.4}{MV/m}               \\
    Straight section length                                       & \SI{4.16}{m}                 \\
    Quadrupole length                                             & \SI{0.4}{m}                  \\
    Quadrupole strength                                           & \SI{\pm 0.21}{T/m}           \\
    Bending section length                                        & \SI{12.5}{m}                 \\
    Bending section circumference                                 & \SI{600}{m}                  \\
    Total circumference                                           & \SI{800}{m}               \\
    Cyclotron frequency                                           & \SI{224}{kHz}                \\
    Revolution time                                               & \SI{4.46}{\micro s}          \\
    $\beta_{x}^{\textnormal{max}},~ \beta_{y}^{\textnormal{max}}$ & \SI{64.54}{m}, \SI{77.39}{m} \\
    Dispersion, $D_{x}^{\textnormal{max}}$                        & \SI{33.81}{m}                \\
    Tunes, $Q_{x}, ~ Q_{y}$                                       & 2.699, 2.245                 \\
    Slip factor, $\frac{dt}{t}/\frac{dp}{p}$                 & -0.253                       \\
    Momentum acceptance, $(dp/p)$                                 & \num{5.2e-4}                 \\
    Horizontal acceptance [\si{mm} \si{mrad}]                               & 4.8                          \\
    RMS emittance [\si{mm} \si{mrad}], $\epsilon_{x}, ~\epsilon_{y}$        & 0.214, 0.250                 \\
    RMS momentum spread                                           & \num{1.177e-4}               \\
    Particles per bunch                                           & \num{1.17e8}                 \\
    RF voltage                                                    & \SI{1.89}{k V}               \\
    Harmonic number, $h$                                          & 80                           \\
    Synchrotron tune, $Q_{s}$                                     & \num{3.81e-3}                \\
    Bucket height, $\Delta p/p_{\textnormal{bucket}}$             & \num{3.77e-4}                \\
    Bucket length                                                 & \SI{10}{m}                   \\
    RMS bunch length, $\sigma_{s}$                                & \SI{0.994}{m}                \\
    Beam planarity                                                & \SI{0.1}{mm}               \\
    CR-beam splitting                                             & \SI{0.01}{mm}                \\
    \bottomrule
  \end{tabular}
  \label{tab:specs}
\end{table}%
\begin{table}[tbp]
  \centering
  \caption{``Magic'' parameters for protons, values obtained from Ref.~\cite{mooser_direct_2014}.}
  \begin{tabular}{lllll}
    \toprule
    $G$           & $\beta$         & $\gamma$        & $p$                  & $KE$
    \\
    \midrule
    $\num{1.793}\qq{}$ & $\num{0.598}\qq{}$ & $\num{1.248}\qq{}$ & $\SI{0.7}{GeV/c}\qq{}$ & $\SI{233}{MeV}$\\
    \bottomrule
  \end{tabular}
  \label{tab:magicgamma}
\end{table}
\section*{Random misalignment of quadrupoles}
\begin{figure}[htbp]
\centering
\includegraphics[width=\textwidth]{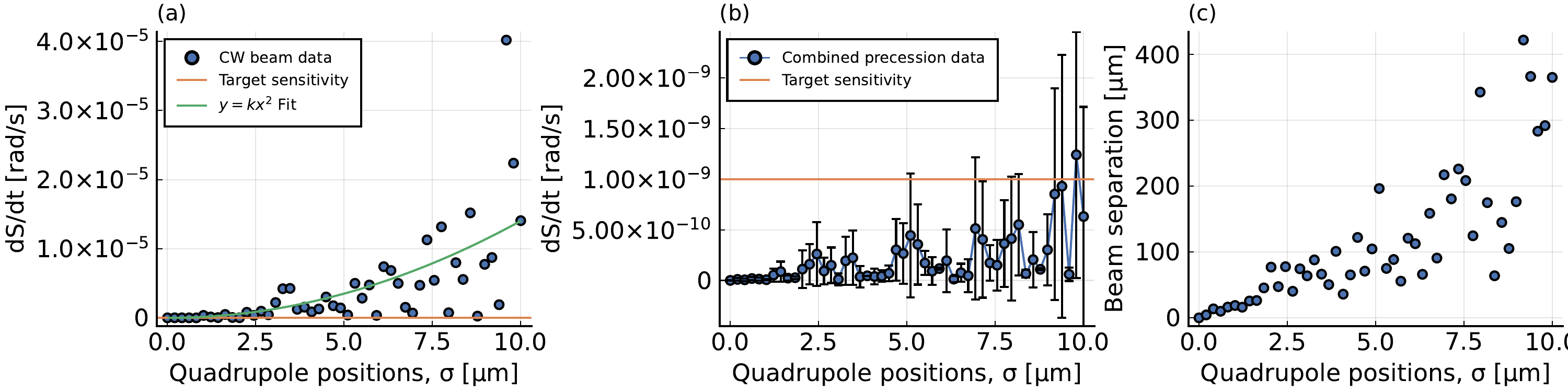}
\caption{\label{fig:combined_geom} Total combined effect of the geometrical
phase. (a) \textit{Longitudinal clockwise.} Absolute value of vertical spin precession rates vs. $\sigma$
quadrupole positions [$\mu$m] in both $x$ and $y$ directions (different random seeds
were used for each point). (b) Same as (a) but complete data combination of CW
and CCW with polarity switching is used. A large cancellation is achieved,
allowing up to $\SI{10}{\micro \metre}$ random quadrupole misalignment. (c) Beam
separation vs. quadrupole positions $\sigma$. As long as the beam separation
can be measured to better than $\SI{100}{\micro m}$, the geometrical phase
should be under control.}
\end{figure}

Random misalignment of quadrupoles in both $x,y$ directions leads to various systematic error sources. The systematic error sources directly caused by it are:
\begin{itemize}
    \item Radial magnetic field.
    \item Vertical magnetic field.
    \item Vertical velocity.
    \item Geometric phase.
\end{itemize}
By randomly moving quadrupoles in the $x,y$ direction by  various $\sigma$ amounts we can estimate the effect of such systematics.
In addition, by repeating this procedure with multiple random seeds, we can eliminate the possibility of a ``lucky configuration'' --- \Cref{fig:combined_geom}.

\section*{Spin Coherence Time}
Spin Coherence Time (SCT), which is also recognized as in-plane polarization (IPP) lifetime, stands for the amount of time that the beam can stay longitudinally
polarized. An SCT of around $\SI{e3}{s}$ is required for the proton EDM
experiment~\cite{edm_proposal}. 

In order to demonstrate a large SCT, sextupoles with strengths $k^m_{1,2}$ are placed
within (on top of) the magnetic quadrupoles. The sextupole fields are defined
as,
\begin{align*}
  B_x &= 2 k^{m} x y \\
  B_y &= k^{m} ( x^2 - y^2 ).
\end{align*}
Effectively, the entire storage ring is now covered with 24 sextupoles of
strength $k^m_1$ and 24 sextupoles of strength $k^m_2$ (following the
alternating pattern as the quadrupoles). In other words, the quadrupoles in
addition to normal operation also act as sextupoles.

Although using correct magnetic sextupoles leads to a prolonged SCT, the same
set of $k^m_{1,2}$ does not lead to a long SCT for both CR beams. A natural attempt 
would be to see how electrical sextupoles
$k^e_{1,2}$ that are similar in strength affect the SCT, where the electric
sextupoles are defined as,
\begin{align*}
  E_x &= -2 k^{e} x y \\
  E_y &= k^{e} ( x^2 - y^2 ).
\end{align*}

If we assign magnetic sextupoles strength $k^m= k^m_1=-k^m_2$ (alternating in sign like magnetic quadrupoles), and electric sextupoles $k^e= k^e_1  =k^e_2$ (same in sign like electrostatic deflectors), CW-CCW symmetry should be conserved in principle. By combining magnetic and electric sextupoles—``hybrid sextupoles''—the equivalence of CW-CCW is restored.

\begin{figure}[tbh]
  \centering
  \includegraphics[width=\textwidth]{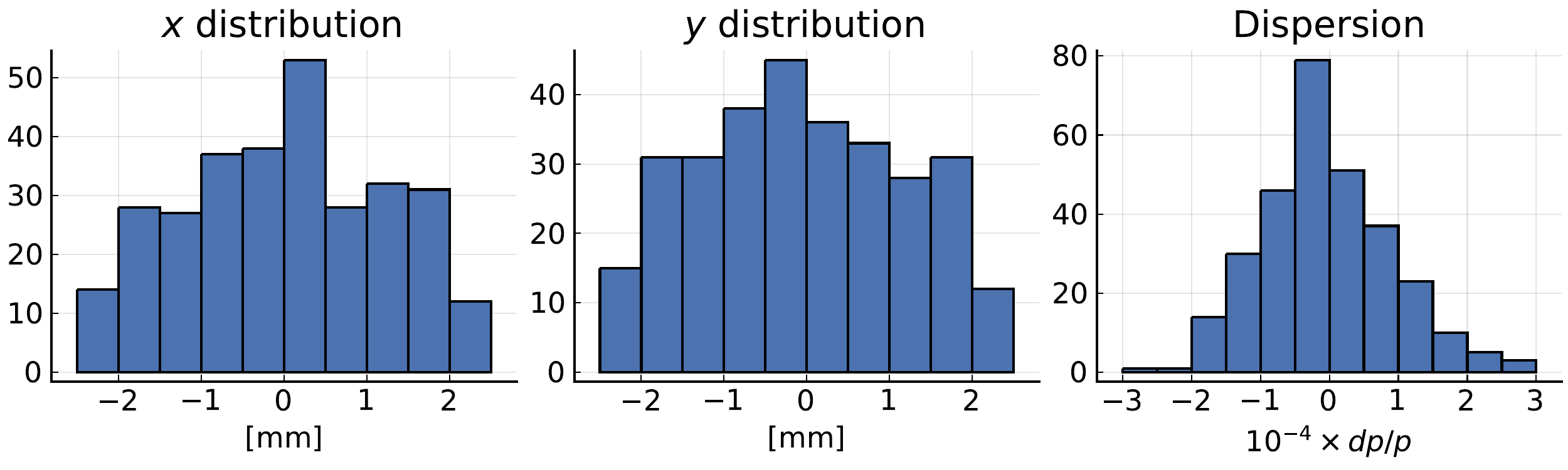}
  \caption{Bunch structure for both CR beams that is used to simulate the polarization lifetime, as shown in
  \Cref{fig:bunch_sct}.}\label{fig:bunch_structure} 
\end{figure}

\begin{figure}[tbh]
  \centering
  \includegraphics[width=\textwidth]{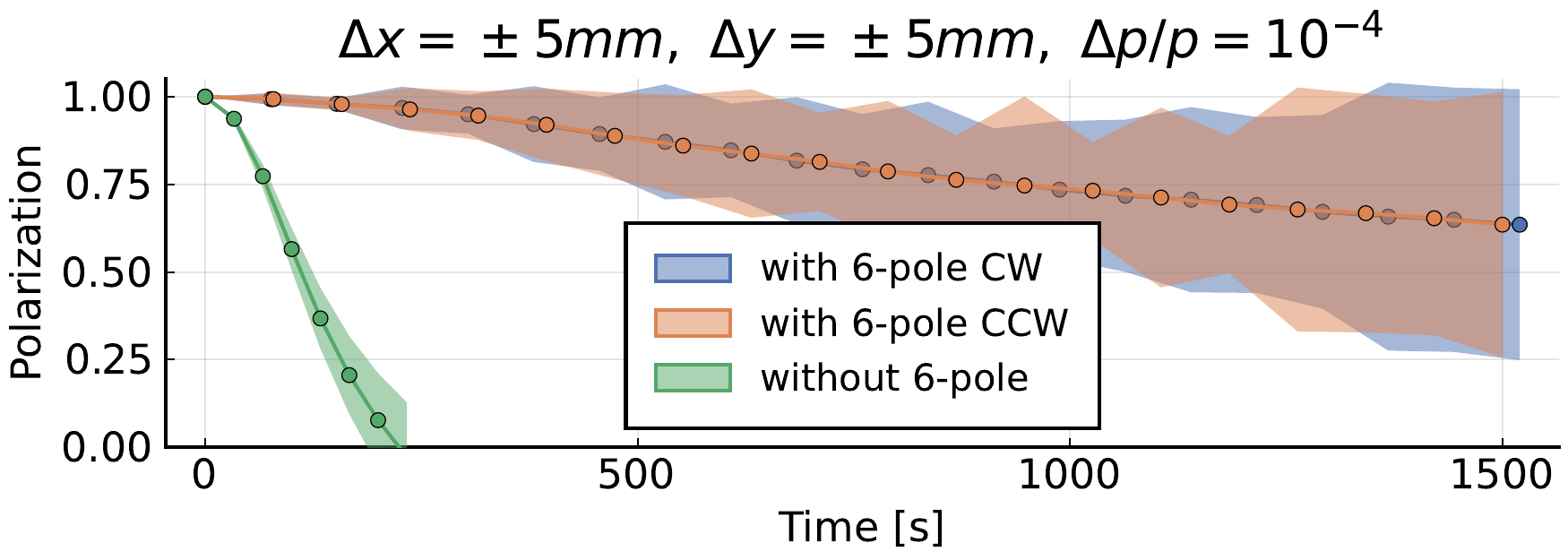}
  \caption{SCT is vastly prolonged when using the correct set of hybrid
  sextupoles.}\label{fig:bunch_sct} 
\end{figure}

By using a realistic bunch structure (\Cref{fig:bunch_structure}), we see 
a large SCT improvement when using the hybrid set of sextupoles ---
\Cref{fig:bunch_sct}.


\section*{Polarimetry}
 
Tests with beams and polarimeters at several laboratories (BNL, KVI, COSY) have consistently demonstrated over more than a decade that the requirements of storage ring EDM search are within reach~\cite{brantjes2012correcting,ref1,ref2,ref3,hempelmann-phase-2017,jedi_collaboration_how_2016,ref6,ref7}. Of particular importance, it has been shown that polarimeters based on forward elastic scattering offer a way to calibrate and correct geometrical and counting rate systematic errors in real time. Sextupole field adjustments along with electron cooling yield long lifetimes for a ring-plane polarization whose direction may be controlled using polarimeter-based feedback. Given the extensive model-based studies demonstrating that ring designs using the symmetries described above can control EDM systematics at the \targetsens\ level~\cite{symmetric}, the optimum path forward is to continue these developments on a full-scale hybrid, symmetric-lattice machine. 
%Both the BNL and FNAL $(g - 2)$ experiments found new systematic effects with improved sensitivities and were able to manage these successfully. We expect to have a similar experience with the EDM search.

\begin{figure}
\centering
\includegraphics[width=0.57\textwidth]{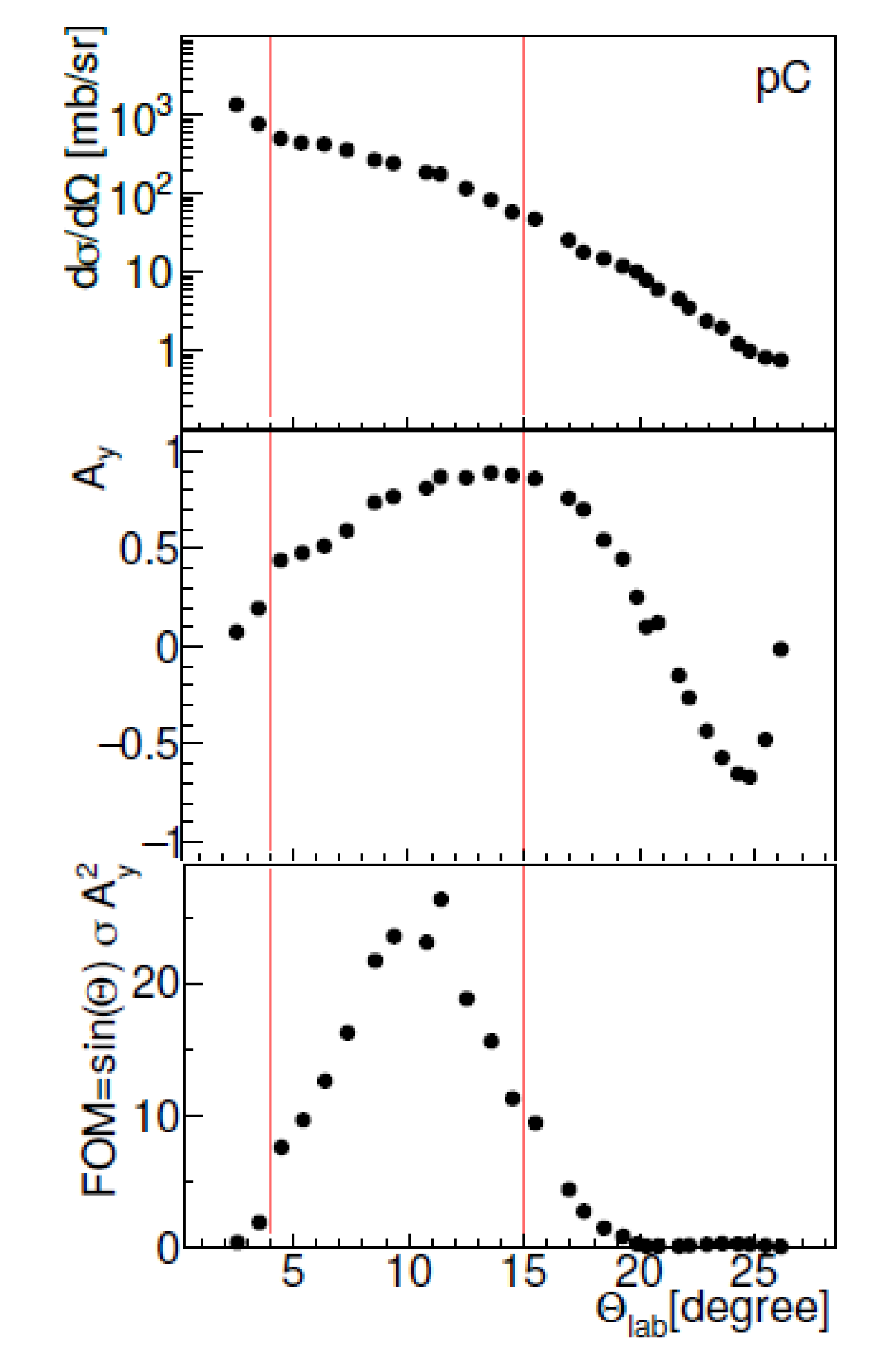}
\caption{Angular distributions of p+C elastic scattering differential cross section, analyzing power, and modified figure of merit ($FOM=(\sin{\theta}) \sigma A^2_y$). The red lines show typical boundaries for data collection in a polarimeter.}\label{fig:FoM}
\end{figure}

The features of the forward-angle elastic scattering polarimeter are listed below:

\begin{itemize}
    \item Carbon target, observing elastic scattering between \ang{5} and \ang{15}. Target thickness: \SIrange{2}{4}{cm}. Angular distributions are shown in \Cref{fig:FoM} from Ref.~\cite{hom88}.
    \item CW and CCW polarimeters share target in middle. Calibrate using vertical polarization.
    \item Detector: position sensitive $\Delta E$, segmented calorimeter.
    \item Efficiency: $\sim 1$\% of the particles removed from beam become part of the useful data stream.
    \item Analyzing power = 0.6, under Monte-Carlo (MC) estimation.
    \item Signal accumulation rate at \targetsens\ is $\SI{e-9}{rad/s}$.
    \item Full azimuthal coverage and forward/backward polarization allow first-order systematic error monitoring by using the four counting rates denoted by left/right detectors and forward/backward polarization. One combination of these rates is polarization insensitive while measuring a first-order driver of systematic errors. Corrections to the signal may be made to second-order in this driver in real time, which appears successful in correcting the signal at levels below $10^{-5}$~\cite{brantjes2012correcting}.
\end{itemize}
 
\section*{EDM Statistics}
The statistical sensitivity of a single measurement, as exemplified by the neutron EDM case, is inversely proportional to the beam polarization, the analyzing power, the spin coherence time (SCT) and the square root of the number of detected events. The advantage of the storage ring method over using neutrons is that high-intensity, highly polarized beams with small values in the relevant phase-space parameters are readily available. As a consequence, it is possible to achieve long SCT with horizontally polarized beams, as was calculated analytically and demonstrated at COSY~\cite{jedi_collaboration_how_2016,hempelmann-phase-2017}.

Under optimized running conditions, where the beam storage duration is for half the SCT, the EDM statistical sensitivity of the method is given by~\cite{kim_new_2021},
\begin{equation}
    \sigma_d = \frac{2.33\hbar}{P_0 A E \sqrt{k N_{cyc} T_{exp}\tau_p}},
\end{equation}
where  $P_0$ ($\sim 0.8$) is the horizontal beam polarization, $A$ ($\sim0.6$) is the asymmetry, $E$ ($\SI{3.3}{MV/m} = \SI{4.4}{MV/m} \times \SI{600}{m}/\SI{800}{m}$) is the average radial electric field integrated around the ring, $k$ (1\%) is the polarimeter detector efficiency, $N_{cyc}$ ($\sim \num{2e10}$) is the stored particles per cycle, $T_{\textnormal{ exp}}$  (\SI{1e8}{s}) is the total duration of the experiment and $\tau_p$ (\SI{2e3}{s}) is the in-plane (horizontal) beam polarization lifetime (equivalent to SCT). The SCT of \SI{2e3}{s}, i.e., an optimum storage time of \SI{e3}{s}, is assumed here in order to achieve a statistical sensitivity at \targetsens~level, while assuming the total experiment duration is \SI{80}{million} seconds (in practice, corresponding to roughly five calendar years). Such a beam storage might require stochastic cooling due to IBS and beam-gas interactions. The estimated SCT of the beam itself (without stochastic cooling) as indicated by preliminary results with high-precision beam/spin-dynamics simulations is greater than \SI{2e3}{s}, limited by the simulation speed. 

\section*{Search for  Axion-like Dark Matter in Storage Rings}
Axion-like dark matter (DM) interacts with a nuclear EDM~\cite{Chadha-Day2022,youn2022}:
\begin{align}
    \mathcal{H} \propto g_\text{EDM} \, a \, \hat{\mathbf{S}} \cdot \mathbf{E}.
\end{align}
This interaction induces an oscillating EDM, since $a$ is a dynamic field: $d_n (t) = g_d a = d_0 \cos(m_a t)$, where $m_a$ is the axion mass. Assuming that it makes up 100\% of the local dark matter, the QCD axion induces an oscillating EDM of approximately $ \num{1e-34}~ e \cdot \textnormal{cm}$~\cite{Graham2013}. Axion-like particles (ALPs), which also may constitute the local DM, are less constrained than those of the QCD axion, motivating experimental searches even above the QCD axion band in the coupling parameter space.

Exploiting the dynamic nature of the nuclear EDM induced by the axion-like DM, proposed experimental approaches aim to enhance the signal using resonances, e.g. nuclear magnetic resonance in the CASPEr experiment~\cite{Budker2014, Kimball2017, Aybas2021} and vertical rotation of the polarization in the storage-ring axion-induced EDM experiment~\cite{chang2019axionlike, kim_new_2021}. The latter is conceptually similar to the storage-ring proton EDM experiment but it does not require the frozen-spin condition.

\begin{figure}
\centering
\includegraphics[width=0.8\textwidth]{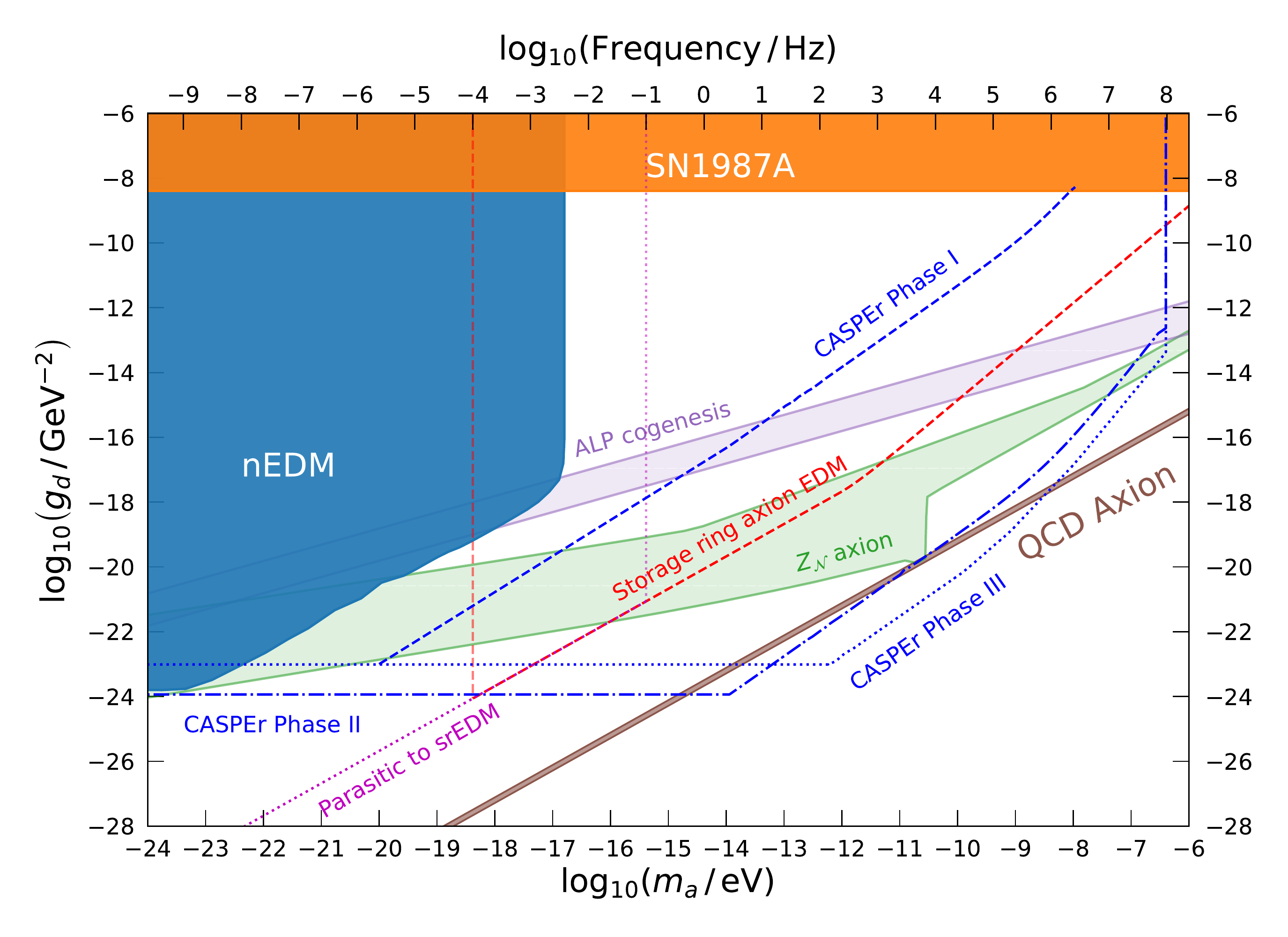}
\caption{Parameter space for the ALP-EDM coupling strength $g_d$. Filled regions are excluded from (blue) the laboratory neutron EDM experiment~\cite{Abel2017} and (orange) astronomic constraints from the supernova cooling~\cite{Graham2013}. The other shaded regions are theoretically motivated regions from (brown) the QCD axion, (purple) the ALP cogenesis when the coupling constant $c_{aNN}$ is between 1 and 10~\cite{Co2021} and (green) $Z_\mathcal{N}$ axion when it makes up the entire local dark matter density~\cite{DiLuzio2021, DiLuzio2021_2}. Dashed lines indicate projected sensitivities proposed by (blue) the CASPEr experiment~\cite{Budker2014, Kimball2017, Aybas2021} and (red) the storage-ring axion-induced EDM experiment including (magenta) parasitic measurement in the frozen-spin storage ring EDM experiment~\cite{chang2019axionlike, kim_new_2021}.} \label{fig:axionEDM_sensitivity}
\end{figure}

Figure~\ref{fig:axionEDM_sensitivity} shows the ALP-EDM coupling parameter space, superimposed by experimentally excluded regions by (blue-filled) the neutron EDM measurement~\cite{Abel2017} and (orange-filled) the supernova energy loss~\cite{Graham2013}; theoretically plausible regions by (brown) the QCD axion band; (purple) ALP cogenesis where its lower and upper bounds correspond to $c_{aNN}=1$ and 10~\cite{Co2021}, respectively; (green) $Z_\mathcal{N}$ axion when it can account for the entire DM density~\cite{DiLuzio2021, DiLuzio2021_2}, and projected experimental sensitivity for (red-dashed) the storage-ring axion-induced EDM experiment including (magenta-dashed) parasitic measurement in the frozen-spin storage ring EDM experiment~\cite{chang2019axionlike, kim_new_2021}; and the CASPEr experiments~\cite{Budker2014, Kimball2017, Aybas2021}. For the storage-ring axion-induced EDM experiment, it assumes a spin coherence time of ${10}^4$ seconds and one year of scientific data accumulation at each frequency, with  100~MV/m effective electric field ($E^* \equiv E - vB$) in the storage ring. There has also appeared a new constraint from the cold neutron-beam experiment at ILL\cite{Schulthess2022}; it has not been included in Fig.~\ref{fig:axionEDM_sensitivity} since it is not published yet.

The storage ring EDM method also allows us to look for ALP-nucleon coupling $\mathcal{H} \propto g_{aNN} \nabla a \cdot \hat{\mathbf{S}}$, as proposed in Ref. \cite{graham_paper}. This interaction also induces a spin precession proportional to $g_{aNN} \cos(m_a t)$. A magnitude of the axion field gradient $\nabla a$ is boosted significantly when filled by a relativistic particle in a storage ring, providing a promising sensitivity on $g_{aNN}$ with dedicated experimental configurations.

\section*{Conclusions}
A storage ring proton EDM experiment offering unprecedented statistical sensitivity to the \targetsens~level can be built based on present technology. The proposed method is based on the hybrid-symmetric ring lattice, the only lattice that eliminates the main EDM systematic error sources within the capacity of present technology.
At the \targetsens~level, this would be the best EDM experiment using one of the simplest hadrons. The facility would also permit studying the deuteron/$^3\textnormal{He}$ EDM with about an order of magnitude lower
sensitivity. Finally, DM/DE experiments running in parasitic mode could probe previously unexplored parameter space.

\pagebreak
\printbibliography

@misc{edmtheory,
title={Overview {EDM} theory},
author={W. Marciano},
journal={Workshop on Electric and Magnetic Dipole Moments, 15 September 2020},
note={\url{https://indico.fnal.gov/event/44782/timetable/?view=nicecompact}}
}

@misc{edmtheory2,
title={{Developing New Directions in Fundamental Physics 2020}},
author={N. Hutzler},
note={\url{https://meetings.triumf.ca/event/89/contributions/2707}}
}

@article{RevModPhys.91.015001,
  title = {Electric dipole moments of atoms, molecules, nuclei, and particles},
  author = {Chupp, T. E. and Fierlinger, P. and Ramsey-Musolf, M. J. and Singh, J. T.},
  journal = {Rev. Mod. Phys.},
  volume = {91},
  issue = {1},
  pages = {015001},
  numpages = {55},
  year = {2019},
  month = {1},
  publisher = {American Physical Society},
  doi = {10.1103/RevModPhys.91.015001},
  url = {https://link.aps.org/doi/10.1103/RevModPhys.91.015001}
}

@article{mooser_direct_2014,
	title = {Direct high-precision measurement of the magnetic moment of the proton},
	volume = {509},
	rights = {2014 Nature Publishing Group, a division of Macmillan Publishers Limited. All Rights Reserved.},
	issn = {1476-4687},
	url = {https://www.nature.com/articles/nature13388},
	doi = {10.1038/nature13388},
	pages = {596--599},
	number = {7502},
	journal = {Nature},
	author = {Mooser, A. and Ulmer, S. and Blaum, K. and Franke, K. and Kracke, H. and Leiteritz, C. and Quint, W. and Rodegheri, C. C. and Smorra, C. and Walz, J.},
	urldate = {2021-03-22},
	year = {2014-05},
	langid = {english},
}

@article{bnl_edm,
  title = {Improved limit on the muon electric dipole moment},
  author = {Bennett, G. W. and Bousquet, B. and Brown, H. N. and Bunce, G. and Carey, R. M. and Cushman, P. and Danby, G. T. and Debevec, P. T. and Deile, M. and Deng, H. and Deninger, W. and Dhawan, S. K. and Druzhinin, V. P. and Duong, L. and Efstathiadis, E. and Farley, F. J. M. and Fedotovich, G. V. and Giron, S. and Gray, F. E. and Grigoriev, D. and Grosse-Perdekamp, M. and Grossmann, A. and Hare, M. F. and Hertzog, D. W. and Huang, X. and Hughes, V. W. and Iwasaki, M. and Jungmann, K. and Kawall, D. and Kawamura, M. and Khazin, B. I. and Kindem, J. and Krienen, F. and Kronkvist, I. and Lam, A. and Larsen, R. and Lee, Y. Y. and Logashenko, I. and McNabb, R. and Meng, W. and Mi, J. and Miller, J. P. and Mizumachi, Y. and Morse, W. M. and Nikas, D. and Onderwater, C. J. G. and Orlov, Y. and \"Ozben, C. S. and Paley, J. M. and Peng, Q. and Polly, C. C. and Pretz, J. and Prigl, R. and zu Putlitz, G. and Qian, T. and Redin, S. I. and Rind, O. and Roberts, B. L. and Ryskulov, N. and Sedykh, S. and Semertzidis, Y. K. and Shagin, P. and Shatunov, Yu. M. and Sichtermann, E. P. and Solodov, E. and Sossong, M. and Steinmetz, A. and Sulak, L. R. and Timmermans, C. and Trofimov, A. and Urner, D. and von Walter, P. and Warburton, D. and Winn, D. and Yamamoto, A. and Zimmerman, D.},
  collaboration = {Muon (g-2) Collaboration},
  journal = {Phys. Rev. D},
  volume = {80},
  issue = {5},
  pages = {052008},
  numpages = {18},
  year = {2009},
  month = {Sep},
  publisher = {American Physical Society},
  doi = {10.1103/PhysRevD.80.052008},
  url = {https://link.aps.org/doi/10.1103/PhysRevD.80.052008}
}

@article{farley_new_2004,
	title = {{New Method of Measuring Electric Dipole Moments in Storage Rings}},
	volume = {93},
	url = {https://link.aps.org/doi/10.1103/PhysRevLett.93.052001},
	doi = {10.1103/PhysRevLett.93.052001},
	pages = {052001},
	number = {5},
	journal = {Physical Review Letters},
	shortjournal = {Phys. Rev. Lett.},
	author = {Farley, F. J. M. and Jungmann, K. and Miller, J. P. and Morse, W. M. and Orlov, Y. F. and Roberts, B. L. and Semertzidis, Y. K. and Silenko, A. and Stephenson, E. J.},
	urldate = {2020-11-17},
	year = {2004-07-27},
}

@article{symmetric,
  title = {Comprehensive symmetric-hybrid ring design for a proton {EDM} experiment at below ${10}^{\ensuremath{-}29}\mathrm{e}\ifmmode\cdot\else\textperiodcentered\fi{}\mathrm{cm}$},
  author = {Omarov, Zhanibek and Davoudiasl, Hooman and Hac{\i}{\"o}mero{\u{g}}lu, Selcuk and Lebedev, Valeri and Morse, William M. and Semertzidis, Yannis K. and Silenko, Alexander J. and Stephenson, Edward J. and Suleiman, Riad},
  journal = {Phys. Rev. D},
  volume = {105},
  issue = {3},
  pages = {032001},
  numpages = {20},
  year = {2022},
  month = {Feb},
  publisher = {American Physical Society},
  doi = {10.1103/PhysRevD.105.032001},
  url = {https://link.aps.org/doi/10.1103/PhysRevD.105.032001}
}

@misc{ags_proposal,
title= {{AGS Proposal: Search for a permanent electric dipole moment of the deuteron nucleus at the $10^{-29}e \cdot$cm level.}},
author = {Anastassopoulos, V. and others},
note = {Access at:  \url{https://www.bnl.gov/edm/files/pdf/deuteron_proposal_080423_final.pdf}},
year = {2008-04}
}

@article{anastassopoulos_storage_2016,
	title = {A storage ring experiment to detect a proton electric dipole moment},
	volume = {87},
	issn = {0034-6748},
	url = {https://aip.scitation.org/doi/10.1063/1.4967465},
	doi = {10.1063/1.4967465},
	pages = {115116},
	number = {11},
	journal = {Review of Scientific Instruments},
	author = {Anastassopoulos, V. and Andrianov, S. and Baartman, R. and Baessler, S. and Bai, M. and Benante, J. and Berz, M. and Blaskiewicz, M. and Bowcock, T. and Brown, K. and Casey, B. and Conte, M. and Crnkovic, J. D. and D’Imperio, N. and Fanourakis, G. and Fedotov, A. and Fierlinger, P. and Fischer, W. and Gaisser, M. O. and Giomataris, Y. and Grosse-Perdekamp, M. and Guidoboni, G. and Hacıömeroğlu, S. and Hoffstaetter, G. and Huang, H. and Incagli, M. and Ivanov, A. and Kawall, D. and Kim, Y. I. and King, B. and Koop, I. A. and Lazarus, D. M. and Lebedev, V. and Lee, M. J. and Lee, S. and Lee, Y. H. and Lehrach, A. and Lenisa, P. and Sandri, P. Levi and Luccio, A. U. and Lyapin, A. and {MacKay}, W. and Maier, R. and Makino, K. and Malitsky, N. and Marciano, W. J. and Meng, W. and Meot, F. and Metodiev, E. M. and Miceli, L. and Moricciani, D. and Morse, W. M. and Nagaitsev, S. and Nayak, S. K. and Orlov, Y. F. and Ozben, C. S. and Park, S. T. and Pesce, A. and Petrakou, E. and Pile, P. and Podobedov, B. and Polychronakos, V. and Pretz, J. and Ptitsyn, V. and Ramberg, E. and Raparia, D. and Rathmann, F. and Rescia, S. and Roser, T. and Sayed, H. Kamal and Semertzidis, Y. K. and Senichev, Y. and Sidorin, A. and Silenko, A. and Simos, N. and Stahl, A. and Stephenson, E. J. and Ströher, H. and Syphers, M. J. and Talman, J. and Talman, R. M. and Tishchenko, V. and Touramanis, C. and Tsoupas, N. and Venanzoni, G. and Vetter, K. and Vlassis, S. and Won, E. and Zavattini, G. and Zelenski, A. and Zioutas, K.},
	urldate = {2018-12-19},
	year = {2016-11-29},
	langid = {english}
}

@article{jedi_collaboration_how_2016,
  title = {{How to Reach a Thousand-Second in-Plane Polarization Lifetime with $0.97\text{\ensuremath{-}}\mathrm{GeV}/c$ Deuterons in a Storage Ring}},
  author = {Guidoboni, G. and Stephenson, E. and Andrianov, S. and Augustyniak, W. and Bagdasarian, Z. and Bai, M. and Baylac, M. and Bernreuther, W. and Bertelli, S. and Berz, M. and B\"oker, J. and B\"ohme, C. and Bsaisou, J. and Chekmenev, S. and Chiladze, D. and Ciullo, G. and Contalbrigo, M. and de Conto, J.-M. and Dymov, S. and Engels, R. and Esser, F. M. and Eversmann, D. and Felden, O. and Gaisser, M. and Gebel, R. and Gl\"uckler, H. and Goldenbaum, F. and Grigoryev, K. and Grzonka, D. and Hahnraths, T. and Heberling, D. and Hejny, V. and Hempelmann, N. and Hetzel, J. and Hinder, F. and Hipple, R. and H\"olscher, D. and Ivanov, A. and Kacharava, A. and Kamerdzhiev, V. and Kamys, B. and Keshelashvili, I. and Khoukaz, A. and Koop, I. and Krause, H.-J. and Krewald, S. and Kulikov, A. and Lehrach, A. and Lenisa, P. and Lomidze, N. and Lorentz, B. and Maanen, P. and Macharashvili, G. and Magiera, A. and Maier, R. and Makino, K. and Maria\ifmmode \acute{n}\else \'{n}\fi{}ski, B. and Mchedlishvili, D. and Mei\ss{}ner, Ulf-G. and Mey, S. and Morse, W. and M\"uller, F. and Nass, A. and Natour, G. and Nikolaev, N. and Nioradze, M. and Nowakowski, K. and Orlov, Y. and Pesce, A. and Prasuhn, D. and Pretz, J. and Rathmann, F. and Ritman, J. and Rosenthal, M. and Rudy, Z. and Saleev, A. and Sefzick, T. and Semertzidis, Y. K. and Senichev, Y. and Shmakova, V. and Silenko, A. and Simon, M. and Slim, J. and Soltner, H. and Stahl, A. and Stassen, R. and Statera, M. and Stockhorst, H. and Straatmann, H. and Str\"oher, H. and Tabidze, M. and Talman, R. and Th\"orngren Engblom, P. and Trinkel, F. and Trzci\ifmmode \acute{n}\else \'{n}\fi{}ski, A. and Uzikov, Yu. and Valdau, Yu. and Valetov, E. and Vassiliev, A. and Weidemann, C. and Wilkin, C. and Wro\ifmmode \acute{n}\else \'{n}\fi{}ska, A. and W\"ustner, P. and Zakrzewska, M. and Zupra\ifmmode \acute{n}\else \'{n}\fi{}ski, P. and Zyuzin, D.},
  collaboration = {JEDI Collaboration},
  journal = {Phys. Rev. Lett.},
  volume = {117},
  issue = {5},
  pages = {054801},
  numpages = {6},
  year = {2016},
  month = {Jul},
  publisher = {American Physical Society},
  doi = {10.1103/PhysRevLett.117.054801},
  url = {https://link.aps.org/doi/10.1103/PhysRevLett.117.054801}
}

@article{hempelmann-phase-2017,
  title = {{Phase Locking the Spin Precession in a Storage Ring}},
  author = {Hempelmann, N. and Hejny, V. and Pretz, J. and Stephenson, E. and Augustyniak, W. and Bagdasarian, Z. and Bai, M. and Barion, L. and Berz, M. and Chekmenev, S. and Ciullo, G. and Dymov, S. and Etzkorn, F.-J. and Eversmann, D. and Gaisser, M. and Gebel, R. and Grigoryev, K. and Grzonka, D. and Guidoboni, G. and Hanraths, T. and Heberling, D. and Hetzel, J. and Hinder, F. and Kacharava, A. and Kamerdzhiev, V. and Keshelashvili, I. and Koop, I. and Kulikov, A. and Lehrach, A. and Lenisa, P. and Lomidze, N. and Lorentz, B. and Maanen, P. and Macharashvili, G. and Magiera, A. and Mchedlishvili, D. and Mey, S. and M\"uller, F. and Nass, A. and Nikolaev, N. N. and Pesce, A. and Prasuhn, D. and Rathmann, F. and Rosenthal, M. and Saleev, A. and Schmidt, V. and Semertzidis, Y. and Shmakova, V. and Silenko, A. and Slim, J. and Soltner, H. and Stahl, A. and Stassen, R. and Stockhorst, H. and Str\"oher, H. and Tabidze, M. and Tagliente, G. and Talman, R. and Th\"orngren Engblom, P. and Trinkel, F. and Uzikov, Yu. and Valdau, Yu. and Valetov, E. and Vassiliev, A. and Weidemann, C. and Wro\ifmmode \acute{n}\else \'{n}\fi{}ska, A. and W\"ustner, P. and Zupra\ifmmode \acute{n}\else \'{n}\fi{}ski, P. and \ifmmode \dot{Z}\else \.{Z}\fi{}urek, M.},
  collaboration = {JEDI Collaboration},
  journal = {Phys. Rev. Lett.},
  volume = {119},
  issue = {1},
  pages = {014801},
  numpages = {5},
  year = {2017},
  month = {Jul},
  publisher = {American Physical Society},
  doi = {10.1103/PhysRevLett.119.014801},
  url = {https://link.aps.org/doi/10.1103/PhysRevLett.119.014801}
}

@article{kim_new_2021,
  title = {New method of probing an oscillating {EDM} induced by axionlike dark matter using an {RF-Wien} filter in storage rings},
  author = {Kim, On and Semertzidis, Yannis K.},
  journal = {Phys. Rev. D},
  volume = {104},
  issue = {9},
  pages = {096006},
  numpages = {12},
  year = {2021},
  month = {Nov},
  publisher = {American Physical Society},
  doi = {10.1103/PhysRevD.104.096006},
  url = {https://link.aps.org/doi/10.1103/PhysRevD.104.096006}
}

@article{SBPM_1,
  title={{SQUID}-based beam position monitor},
  author={Hac{\i}{\"o}mero{\u{g}}lu, Sel{\c{c}}uk  and Kawall, David and Lee, Yong-Ho and Matlashov, Andrei and Omarov, Zhanibek and Semertzidis, Yannis K.},
  journal={The 39th International Conference on High Energy Physics, ICHEP2018},
  year={2018}
}

@article{ref1,
  title = {Phase measurement for driven spin oscillations in a storage ring},
  author = {Hempelmann, N. and Hejny, V. and Pretz, J. and Soltner, H. and Augustyniak, W. and Bagdasarian, Z. and Bai, M. and Barion, L. and Berz, M. and Chekmenev, S. and Ciullo, G. and Dymov, S. and Eversmann, D. and Gaisser, M. and Gebel, R. and Grigoryev, K. and Grzonka, D. and Guidoboni, G. and Heberling, D. and Hetzel, J. and Hinder, F. and Kacharava, A. and Kamerdzhiev, V. and Keshelashvili, I. and Koop, I. and Kulikov, A. and Lehrach, A. and Lenisa, P. and Lomidze, N. and Lorentz, B. and Maanen, P. and Macharashvili, G. and Magiera, A. and Mchedlishvili, D. and Mey, S. and M\"uller, F. and Nass, A. and Nikolaev, N. N. and Nioradze, M. and Pesce, A. and Prasuhn, D. and Rathmann, F. and Rosenthal, M. and Saleev, A. and Schmidt, V. and Semertzidis, Y. and Senichev, Y. and Shmakova, V. and Silenko, A. and Slim, J. and Stahl, A. and Stassen, R. and Stephenson, E. and Stockhorst, H. and Str\"oher, H. and Tabidze, M. and Tagliente, G. and Talman, R. and Th\"orngren Engblom, P. and Trinkel, F. and Uzikov, Yu. and Valdau, Yu. and Valetov, E. and Vassiliev, A. and Weidemann, C. and Wro\ifmmode \acute{n}\else \'{n}\fi{}ska, A. and W\"ustner, P. and Zupra\ifmmode \acute{n}\else \'{n}\fi{}ski, P. and \ifmmode \dot{Z}\else \.{Z}\fi{}urek, M.},
  collaboration = {JEDI Collaboration},
  journal = {Phys. Rev. Accel. Beams},
  volume = {21},
  issue = {4},
  pages = {042002},
  numpages = {6},
  year = {2018},
  month = {Apr},
  publisher = {American Physical Society},
  doi = {10.1103/PhysRevAccelBeams.21.042002},
  url = {https://link.aps.org/doi/10.1103/PhysRevAccelBeams.21.042002}
}

@article{ref2,
  title = {Connection between zero chromaticity and long in-plane polarization lifetime in a magnetic storage ring},
  author = {Guidoboni, G. and Stephenson, E. J. and Wro\ifmmode \acute{n}\else \'{n}\fi{}ska, A. and Bagdasarian, Z. and Bsaisou, J. and Chekmenev, S. and Ciullo, G. and Dymov, S. and Eversmann, D. and Gaisser, M. and Gebel, R. and Hejny, V. and Hempelmann, N. and Hinder, F. and Kacharava, A. and Keshelashvili, I. and Kulessa, P. and Lenisa, P. and Lehrach, A. and Lorentz, B. and Maanen, P. and Maier, R. and Mchedlishvili, D. and Mey, S. and Nass, A. and Pesce, A. and Orlov, Y. and Pretz, J. and Prasuhn, D. and Rathmann, F. and Rosenthal, M. and Saleev, A. and Semertzidis, Y. K. and Senichev, Y. and Shmakova, V. and Stockhorst, H. and Str\"oher, H. and Talman, R. and Th\"orngren Engblom, P. and Trinkel, F. and Valdau, Yu. and Weidemann, C. and W\"ustner, P. and \ifmmode \dot{Z}\else \.{Z}\fi{}urek, M. and Zyuzin, D.},
  collaboration = {JEDI Collaboration},
  journal = {Phys. Rev. Accel. Beams},
  volume = {21},
  issue = {2},
  pages = {024201},
  numpages = {14},
  year = {2018},
  month = {Feb},
  publisher = {American Physical Society},
  doi = {10.1103/PhysRevAccelBeams.21.024201},
  url = {https://link.aps.org/doi/10.1103/PhysRevAccelBeams.21.024201}
}

@article{ref3,
  title = {Spin tune mapping as a novel tool to probe the spin dynamics in storage rings},
  author = {Saleev, A. and Nikolaev, N. N. and Rathmann, F. and Augustyniak, W. and Bagdasarian, Z. and Bai, M. and Barion, L. and Berz, M. and Chekmenev, S. and Ciullo, G. and Dymov, S. and Eversmann, D. and Gaisser, M. and Gebel, R. and Grigoryev, K. and Grzonka, D. and Guidoboni, G. and Heberling, D. and Hejny, V. and Hempelmann, N. and Hetzel, J. and Hinder, F. and Kacharava, A. and Kamerdzhiev, V. and Keshelashvili, I. and Koop, I. and Kulikov, A. and Lehrach, A. and Lenisa, P. and Lomidze, N. and Lorentz, B. and Maanen, P. and Macharashvili, G. and Magiera, A. and Mchedlishvili, D. and Mey, S. and M\"uller, F. and Nass, A. and Pesce, A. and Prasuhn, D. and Pretz, J. and Rosenthal, M. and Schmidt, V. and Semertzidis, Y. and Senichev, Y. and Shmakova, V. and Silenko, A. and Slim, J. and Soltner, H. and Stahl, A. and Stassen, R. and Stephenson, E. and Stockhorst, H. and Str\"oher, H. and Tabidze, M. and Tagliente, G. and Talman, R. and Engblom, P. Th\"orngren and Trinkel, F. and Uzikov, Yu. and Valdau, Yu. and Valetov, E. and Vassiliev, A. and Weidemann, C. and Wro\ifmmode \acute{n}\else \'{n}\fi{}ska, A. and W\"ustner, P. and Zupra\ifmmode \acute{n}\else \'{n}\fi{}ski, P. and Zurek, M.},
  collaboration = {JEDI collaboration},
  journal = {Phys. Rev. Accel. Beams},
  volume = {20},
  issue = {7},
  pages = {072801},
  numpages = {31},
  year = {2017},
  month = {Jul},
  publisher = {American Physical Society},
  doi = {10.1103/PhysRevAccelBeams.20.072801},
  url = {https://link.aps.org/doi/10.1103/PhysRevAccelBeams.20.072801}
}

@article{ref6,
  title = {{New Method for a Continuous Determination of the Spin Tune in Storage Rings and Implications for Precision Experiments}},
  author = {Eversmann, D. and Hejny, V. and Hinder, F. and Kacharava, A. and Pretz, J. and Rathmann, F. and Rosenthal, M. and Trinkel, F. and Andrianov, S. and Augustyniak, W. and Bagdasarian, Z. and Bai, M. and Bernreuther, W. and Bertelli, S. and Berz, M. and Bsaisou, J. and Chekmenev, S. and Chiladze, D. and Ciullo, G. and Contalbrigo, M. and de Vries, J. and Dymov, S. and Engels, R. and Esser, F. M. and Felden, O. and Gaisser, M. and Gebel, R. and Gl\"uckler, H. and Goldenbaum, F. and Grigoryev, K. and Grzonka, D. and Guidoboni, G. and Hanhart, C. and Heberling, D. and Hempelmann, N. and Hetzel, J. and Hipple, R. and H\"olscher, D. and Ivanov, A. and Kamerdzhiev, V. and Kamys, B. and Keshelashvili, I. and Khoukaz, A. and Koop, I. and Krause, H.-J. and Krewald, S. and Kulikov, A. and Lehrach, A. and Lenisa, P. and Lomidze, N. and Lorentz, B. and Maanen, P. and Macharashvili, G. and Magiera, A. and Maier, R. and Makino, K. and Maria\ifmmode \acute{n}\else \'{n}\fi{}ski, B. and Mchedlishvili, D. and Mei\ss{}ner, Ulf-G. and Mey, S. and Nass, A. and Natour, G. and Nikolaev, N. and Nioradze, M. and Nogga, A. and Nowakowski, K. and Pesce, A. and Prasuhn, D. and Ritman, J. and Rudy, Z. and Saleev, A. and Semertzidis, Y. and Senichev, Y. and Shmakova, V. and Silenko, A. and Slim, J. and Soltner, H. and Stahl, A. and Stassen, R. and Statera, M. and Stephenson, E. and Stockhorst, H. and Straatmann, H. and Str\"oher, H. and Tabidze, M. and Talman, R. and Th\"orngren Engblom, P. and Trzci\ifmmode \acute{n}\else \'{n}\fi{}ski, A. and Uzikov, Yu. and Valdau, Yu. and Valetov, E. and Vassiliev, A. and Weidemann, C. and Wilkin, C. and Wirzba, A. and Wro\ifmmode \acute{n}\else \'{n}\fi{}ska, A. and W\"ustner, P. and Zakrzewska, M. and Zupra\ifmmode \acute{n}\else \'{n}\fi{}ski, P. and Zyuzin, D.},
  collaboration = {JEDI collaboration},
  journal = {Phys. Rev. Lett.},
  volume = {115},
  issue = {9},
  pages = {094801},
  numpages = {6},
  year = {2015},
  month = {Aug},
  publisher = {American Physical Society},
  doi = {10.1103/PhysRevLett.115.094801},
  url = {https://link.aps.org/doi/10.1103/PhysRevLett.115.094801}
}

@article{ref7,
  title = {Measuring the polarization of a rapidly precessing deuteron beam},
  author = {Bagdasarian, Z. and Bertelli, S. and Chiladze, D. and Ciullo, G. and Dietrich, J. and Dymov, S. and Eversmann, D. and Fanourakis, G. and Gaisser, M. and Gebel, R. and Gou, B. and Guidoboni, G. and Hejny, V. and Kacharava, A. and Kamerdzhiev, V. and Lehrach, A. and Lenisa, P. and Lorentz, B. and Magallanes, L. and Maier, R. and Mchedlishvili, D. and Morse, W. M. and Nass, A. and Oellers, D. and Pesce, A. and Prasuhn, D. and Pretz, J. and Rathmann, F. and Shmakova, V. and Semertzidis, Y. K. and Stephenson, E. J. and Stockhorst, H. and Str\"oher, H. and Talman, R. and Th\"orngren Engblom, P. and Valdau, Yu. and Weidemann, C. and W\"ustner, P.},
  journal = {Phys. Rev. ST Accel. Beams},
  volume = {17},
  issue = {5},
  pages = {052803},
  numpages = {15},
  year = {2014},
  month = {May},
  publisher = {American Physical Society},
  doi = {10.1103/PhysRevSTAB.17.052803},
  url = {https://link.aps.org/doi/10.1103/PhysRevSTAB.17.052803}
}

@article{brantjes2012correcting,
  title={Correcting systematic errors in high-sensitivity deuteron polarization measurements},
  author={Brantjes, NPM and Dzordzhadze, V and Gebel, R and Gonnella, F and Gray, FE and Van Der Hoek, DJ and Imig, A and Kruithof, WL and Lazarus, DM and Lehrach, A and others},
  journal={Nuclear Instruments and Methods in Physics Research Section A: Accelerators, Spectrometers, Detectors and Associated Equipment},
  volume={664},
  number={1},
  pages={49--64},
  year={2012},
  publisher={Elsevier}
}

@article{metodiev_fringe_2014,
	title = {Fringe electric fields of flat and cylindrical deflectors in electrostatic charged particle storage rings},
	volume = {17},
	url = {https://link.aps.org/doi/10.1103/PhysRevSTAB.17.074002},
	doi = {10.1103/PhysRevSTAB.17.074002},
	pages = {074002},
	number = {7},
	journal = {Physical Review Special Topics - Accelerators and Beams},
	shortjournal = {Phys. Rev. {ST} Accel. Beams},
	author = {Metodiev, E. M. and Huang, K. L. and Semertzidis, Y. K. and Morse, W. M.},
	urldate = {2020-11-17},
	year = {2014-07-18},
	note = {Publisher: American Physical Society}
}

@article{metodiev_analytical_2015,
	title = {Analytical benchmarks for precision particle tracking in electric and magnetic rings},
	volume = {797},
	issn = {01689002},
	url = {http://linkinghub.elsevier.com/retrieve/pii/S0168900215007822},
	doi = {10.1016/j.nima.2015.06.032},
	pages = {311--318},
	journal = {Nuclear Instruments and Methods in Physics Research Section A: Accelerators, Spectrometers, Detectors and Associated Equipment},
	author = {Metodiev, E. M. and D'Silva, I. M. and Fandaros, M. and Gaisser, M. and Hacıömeroğlu, S. and Huang, D. and Huang, K. L. and Patil, A. and Prodromou, R. and Semertzidis, O. A. and Sharma, D. and Stamatakis, A. N. and Orlov, Y. F. and Semertzidis, Y. K.},
	urldate = {2018-02-14},
	year = {2015-10},
	langid = {english}
}

@article{morse2013rf,
  title={{RF-Wien} filter in an electric dipole moment storage ring: The `partially frozen spin' effect},
  author={Morse, William M and Orlov, Yuri F and Semertzidis, Yannis K},
  journal={Physical Review Special Topics-Accelerators and Beams},
  volume={16},
  number={11},
  pages={114001},
  year={2013},
  publisher={APS}
}

@article{chang2019axionlike,
  title={{Axionlike dark matter search using the storage ring EDM method}},
  author={Chang, Seung Pyo and Hac{\i}{\"o}mero{\u{g}}lu, Sel{\c{c}}uk and Kim, On and Lee, Soohyung and Park, Seongtae and Semertzidis, Yannis K},
  journal={Physical Review D},
  volume={99},
  number={8},
  pages={083002},
  year={2019},
  publisher={APS}
}

@article{haciomeroglu_hybrid_2018,
  title = {Hybrid ring design in the storage-ring proton electric dipole moment experiment},
  author = {Hac{\i}{\"o}mero{\u{g}}lu, Sel{\c{c}}uk  and Semertzidis, Yannis K},
  journal = {Phys. Rev. Accel. Beams},
  volume = {22},
  issue = {3},
  pages = {034001},
  numpages = {7},
  year = {2019},
  month = {Mar},
  publisher = {American Physical Society},
  doi = {10.1103/PhysRevAccelBeams.22.034001},
  url = {https://link.aps.org/doi/10.1103/PhysRevAccelBeams.22.034001}
}

@article{graham_paper,
  title = {Storage ring probes of dark matter and dark energy},
  author = {Graham, Peter W. and Hac{\i}{\"o}mero{\u{g}}lu, Sel{\c{c}}uk  and Kaplan, David E. and Omarov, Zhanibek and Rajendran, Surjeet and Semertzidis, Yannis K.},
  journal = {Phys. Rev. D},
  volume = {103},
  issue = {5},
  pages = {055010},
  numpages = {13},
  year = {2021},
  month = {Mar},
  publisher = {American Physical Society},
  doi = {10.1103/PhysRevD.103.055010},
  url = {https://link.aps.org/doi/10.1103/PhysRevD.103.055010}
}

@article{hom88,
  title = {{Proton elastic scattering from $^{12}\mathrm{C}$ at 250~MeV and energy dependent potentials between 200 and 300~MeV}},
  author = {Meyer, H. O. and Schwandt, P. and Abegg, R. and Miller, C. A. and Jackson, K. P. and Yen, S. and Gaillard, G. and Hugi, M. and Helmer, R. and Frekers, D. and Saxena, A.},
  journal = {Phys. Rev. C},
  volume = {37},
  issue = {2},
  pages = {544--548},
  numpages = {0},
  year = {1988},
  month = {Feb},
  publisher = {American Physical Society},
  doi = {10.1103/PhysRevC.37.544},
  url = {https://link.aps.org/doi/10.1103/PhysRevC.37.544}
}

@article{bmt3,
  title={Equation of spin motion in storage rings in the cylindrical coordinate system},
  author={Silenko, Alexander J},
  journal={Physical Review Special Topics-Accelerators and Beams},
  volume={9},
  number={3},
  pages={034003},
  year={2006},
  publisher={APS}
}

@article{electrode1,
  title = {Evaluation of niobium as candidate electrode material for dc high voltage photoelectron guns},
  author = {BastaniNejad, M. and Mohamed, Md. Abdullah and Elmustafa, A. A. and Adderley, P. and Clark, J. and Covert, S. and Hansknecht, J. and Hernandez-Garcia, C. and Poelker, M. and Mammei, R. and Surles-Law, K. and Williams, P.},
  journal = {Phys. Rev. ST Accel. Beams},
  volume = {15},
  issue = {8},
  pages = {083502},
  numpages = {11},
  year = {2012},
  month = {Aug},
  publisher = {American Physical Society},
  doi = {10.1103/PhysRevSTAB.15.083502},
  url = {https://link.aps.org/doi/10.1103/PhysRevSTAB.15.083502}
}

@article{electrode2,
  title={{TiN} coated aluminum electrodes for {DC} high voltage electron guns},
  author={Mamun, Md Abdullah A and Elmustafa, Abdelmageed A and Taus, Rhys and Forman, Eric and Poelker, Matthew},
  journal={Journal of Vacuum Science \& Technology A: Vacuum, Surfaces, and Films},
  volume={33},
  number={3},
  pages={031604},
  year={2015},
  publisher={American Vacuum Society}
}

@article{electrode3,
  title={Electrostatic design and conditioning of a triple point junction shield for a -200~{kV DC} high voltage photogun},
  author={Palacios-Serrano, G and Hannon, F and Hernandez-Garcia, Carlos and Poelker, Matt and Baumgart, Helmut},
  journal={Review of Scientific Instruments},
  volume={89},
  number={10},
  pages={104703},
  year={2018},
  publisher={AIP Publishing LLC}
}

@article{fnal1,
  title = {{Measurement of the Positive Muon Anomalous Magnetic Moment to 0.46 ppm}},
  author = {Abi, B. and Albahri, T. and Al-Kilani, S. and Allspach, D. and Alonzi, L. P. and Anastasi, A. and Anisenkov, A. and Azfar, F. and Badgley, K. and Bae\ss{}ler, S. and Bailey, I. and Baranov, V. A. and Barlas-Yucel, E. and Barrett, T. and Barzi, E. and Basti, A. and Bedeschi, F. and Behnke, A. and Berz, M. and Bhattacharya, M. and Binney, H. P. and Bjorkquist, R. and Bloom, P. and Bono, J. and Bottalico, E. and Bowcock, T. and Boyden, D. and Cantatore, G. and Carey, R. M. and Carroll, J. and Casey, B. C. K. and Cauz, D. and Ceravolo, S. and Chakraborty, R. and Chang, S. P. and Chapelain, A. and Chappa, S. and Charity, S. and Chislett, R. and Choi, J. and Chu, Z. and Chupp, T. E. and Convery, M. E. and Conway, A. and Corradi, G. and Corrodi, S. and Cotrozzi, L. and Crnkovic, J. D. and Dabagov, S. and De Lurgio, P. M. and Debevec, P. T. and Di Falco, S. and Di Meo, P. and Di Sciascio, G. and Di Stefano, R. and Drendel, B. and Driutti, A. and Duginov, V. N. and Eads, M. and Eggert, N. and Epps, A. and Esquivel, J. and Farooq, M. and Fatemi, R. and Ferrari, C. and Fertl, M. and Fiedler, A. and Fienberg, A. T. and Fioretti, A. and Flay, D. and Foster, S. B. and Friedsam, H. and Frle\ifmmode \check{z}\else \v{z}\fi{}, E. and Froemming, N. S. and Fry, J. and Fu, C. and Gabbanini, C. and Galati, M. D. and Ganguly, S. and Garcia, A. and Gastler, D. E. and George, J. and Gibbons, L. K. and Gioiosa, A. and Giovanetti, K. L. and Girotti, P. and Gohn, W. and Gorringe, T. and Grange, J. and Grant, S. and Gray, F. and Haciomeroglu, S. and Hahn, D. and Halewood-Leagas, T. and Hampai, D. and Han, F. and Hazen, E. and Hempstead, J. and Henry, S. and Herrod, A. T. and Hertzog, D. W. and Hesketh, G. and Hibbert, A. and Hodge, Z. and Holzbauer, J. L. and Hong, K. W. and Hong, R. and Iacovacci, M. and Incagli, M. and Johnstone, C. and Johnstone, J. A. and Kammel, P. and Kargiantoulakis, M. and Karuza, M. and Kaspar, J. and Kawall, D. and Kelton, L. and Keshavarzi, A. and Kessler, D. and Khaw, K. S. and Khechadoorian, Z. and Khomutov, N. V. and Kiburg, B. and Kiburg, M. and Kim, O. and Kim, S. C. and Kim, Y. I. and King, B. and Kinnaird, N. and Korostelev, M. and Kourbanis, I. and Kraegeloh, E. and Krylov, V. A. and Kuchibhotla, A. and Kuchinskiy, N. A. and Labe, K. R. and LaBounty, J. and Lancaster, M. and Lee, M. J. and Lee, S. and Leo, S. and Li, B. and Li, D. and Li, L. and Logashenko, I. and Lorente Campos, A. and Luc\`a, A. and Lukicov, G. and Luo, G. and Lusiani, A. and Lyon, A. L. and MacCoy, B. and Madrak, R. and Makino, K. and Marignetti, F. and Mastroianni, S. and Maxfield, S. and McEvoy, M. and Merritt, W. and Mikhailichenko, A. A. and Miller, J. P. and Miozzi, S. and Morgan, J. P. and Morse, W. M. and Mott, J. and Motuk, E. and Nath, A. and Newton, D. and Nguyen, H. and Oberling, M. and Osofsky, R. and Ostiguy, J.-F. and Park, S. and Pauletta, G. and Piacentino, G. M. and Pilato, R. N. and Pitts, K. T. and Plaster, B. and Po\ifmmode \check{c}\else \v{c}\fi{}ani\ifmmode \acute{c}\else \'{c}\fi{}, D. and Pohlman, N. and Polly, C. C. and Popovic, M. and Price, J. and Quinn, B. and Raha, N. and Ramachandran, S. and Ramberg, E. and Rider, N. T. and Ritchie, J. L. and Roberts, B. L. and Rubin, D. L. and Santi, L. and Sathyan, D. and Schellman, H. and Schlesier, C. and Schreckenberger, A. and Semertzidis, Y. K. and Shatunov, Y. M. and Shemyakin, D. and Shenk, M. and Sim, D. and Smith, M. W. and Smith, A. and Soha, A. K. and Sorbara, M. and St\"ockinger, D. and Stapleton, J. and Still, D. and Stoughton, C. and Stratakis, D. and Strohman, C. and Stuttard, T. and Swanson, H. E. and Sweetmore, G. and Sweigart, D. A. and Syphers, M. J. and Tarazona, D. A. and Teubner, T. and Tewsley-Booth, A. E. and Thomson, K. and Tishchenko, V. and Tran, N. H. and Turner, W. and Valetov, E. and Vasilkova, D. and Venanzoni, G. and Volnykh, V. P. and Walton, T. and Warren, M. and Weisskopf, A. and Welty-Rieger, L. and Whitley, M. and Winter, P. and Wolski, A. and Wormald, M. and Wu, W. and Yoshikawa, C.},
  collaboration = {Muon $g\ensuremath{-}2$ Collaboration},
  journal = {Phys. Rev. Lett.},
  volume = {126},
  issue = {14},
  pages = {141801},
  numpages = {11},
  year = {2021},
  month = {Apr},
  publisher = {American Physical Society},
  doi = {10.1103/PhysRevLett.126.141801},
  url = {https://link.aps.org/doi/10.1103/PhysRevLett.126.141801}
}

@article{cern3,
  title={{Final report on the CERN muon storage ring including the anomalous magnetic moment and the electric dipole moment of the muon, and a direct test of relativistic time dilation}},
  author={Bailey, J and Borer, K and Combley, F and Drumm, H and Eck, C and Farley, FJM and Field, JH and Flegel, W and Hattersley, PM and Krienen, F and others},
  journal={Nuclear Physics B},
  volume={150},
  url = {https://doi.org/10.1016/0550-3213(79)90292-X},
  doi = {10.1016/0550-3213(79)90292-X},
  pages={1--75},
  year={1979},
  publisher={Elsevier}
}

@article{bennett_final_2006,
	title = {{Final report of the E821 muon anomalous magnetic moment measurement at BNL}},
	volume = {73},
	issn = {1550-7998, 1550-2368},
	url = {https://link.aps.org/doi/10.1103/PhysRevD.73.072003},
	doi = {10.1103/PhysRevD.73.072003},
	number = {7},
	journal = {Physical Review D},
	author = {Bennett, G. W. and Bousquet, B. and Brown, H. N. and Bunce, G. and Carey, R. M. and Cushman, P. and Danby, G. T. and Debevec, P. T. and Deile, M. and Deng, H. and Deninger, W. and Dhawan, S. K. and Druzhinin, V. P. and Duong, L. and Efstathiadis, E. and Farley, F. J. M. and Fedotovich, G. V. and Giron, S. and Gray, F. E. and Grigoriev, D. and Grosse-Perdekamp, M. and Grossmann, A. and Hare, M. F. and Hertzog, D. W. and Huang, X. and Hughes, V. W. and Iwasaki, M. and Jungmann, K. and Kawall, D. and Kawamura, M. and Khazin, B. I. and Kindem, J. and Krienen, F. and Kronkvist, I. and Lam, A. and Larsen, R. and Lee, Y. Y. and Logashenko, I. and {McNabb}, R. and Meng, W. and Mi, J. and Miller, J. P. and Mizumachi, Y. and Morse, W. M. and Nikas, D. and Onderwater, C. J. G. and Orlov, Y. and Özben, C. S. and Paley, J. M. and Peng, Q. and Polly, C. C. and Pretz, J. and Prigl, R. and zu Putlitz, G. and Qian, T. and Redin, S. I. and Rind, O. and Roberts, B. L. and Ryskulov, N. and Sedykh, S. and Semertzidis, Y. K. and Shagin, P. and Shatunov, Yu. M. and Sichtermann, E. P. and Solodov, E. and Sossong, M. and Steinmetz, A. and Sulak, L. R. and Timmermans, C. and Trofimov, A. and Urner, D. and von Walter, P. and Warburton, D. and Winn, D. and Yamamoto, A. and Zimmerman, D.},
	urldate = {2018-02-14},
	year = {2006-04-07},
	langid = {english}
}

@article{cern_report,
  title={{The CERN muon $(g-2)$ experiments}},
  author={Combley, F  and Farley, FJM and Picasso, E.},
  journal={Physics Reports},
  volume={68},
  issn = {0370-1573},
  url = {https://doi.org/10.1016/0370-1573(81)90028-4},
	doi = {10.1016/0370-1573(81)90028-4},
  pages={93--119},
  year={1981},
  publisher={Elsevier}
}

@article{kuchler_2019,
  title={{Searches for Electric Dipole Moments—Overview of Status and New Experimental Efforts}},
  author={Kuchler, Florian},
  collaboration = {and on behalf of TUCAN and HeXeEDM Collaborations},
  journal={Universe},
  volume={5},
  issn = {2218-1997},
  url = {https://www.mdpi.com/2218-1997/5/2/56},
	doi = {10.3390/universe5020056},
  pages={1--11},
  year={2019}
}

@article{golub75,
  title={Super-thermal sources of ultra-cold neutrons},
  author={R. Golub and J.M. Pendlebury},
  journal={Physics Letters A},
  volume={53},
  issn = {0375-9601},
  url = {https://doi.org/10.1016/0375-9601(75)90500-9},
	doi = {10.1016/0375-9601(75)90500-9},
  pages={133-135},
  year={1975},
  publisher={Elsevier}
}

@article{golub77,
  title={{The interaction of Ultra-Cold Neutrons ({UCN}) with liquid helium and a superthermal {UCN} source}},
  author={R. Golub and J.M. Pendlebury},
  journal={Physics Letters A},
  volume={62},
  issn = {0375-9601},
  url = {https://www.sciencedirect.com/science/article/pii/0375960177904340},
	doi = {10.1016/0375-9601(77)90434-0},
  pages={337-339},
  year={1977},
  publisher={Elsevier}
}

@article{shin:2018emy,
    author = "Shin, Yun Chang and Snow, W. Michael and Baxter, David V. and Liu, Chen-Yu and Kim, Dongok and Kim, Younggeun and Semertzidis, Yannis K.",
    title = "{Compact ultracold neutron source concept for low-energy accelerator-driven neutron sources}",
    eprint = "1810.08722",
    archivePrefix = "arXiv",
    primaryClass = "physics.ins-det",
    doi = "10.1140/epjp/s13360-021-01740-1",
    journal = "Eur. Phys. J. Plus",
    volume = "136",
    number = "8",
    pages = "882",
    year = "2021"
}

@article{neutronEDM2020,
  title = {{Measurement of the Permanent Electric Dipole Moment of the Neutron}},
  author = {Abel, C. and Afach, S. and Ayres, N. J. and Baker, C. A. and Ban, G. and Bison, G. and Bodek, K. and Bondar, V. and Burghoff, M. and Chanel, E. and Chowdhuri, Z. and Chiu, P.-J. and Clement, B. and Crawford, C. B. and Daum, M. and Emmenegger, S. and Ferraris-Bouchez, L. and Fertl, M. and Flaux, P. and Franke, B. and Fratangelo, A. and Geltenbort, P. and Green, K. and Griffith, W. C. and van der Grinten, M. and Gruji\ifmmode \acute{c}\else \'{c}\fi{}, Z. D. and Harris, P. G. and Hayen, L. and Heil, W. and Henneck, R. and H\'elaine, V. and Hild, N. and Hodge, Z. and Horras, M. and Iaydjiev, P. and Ivanov, S. N. and Kasprzak, M. and Kermaidic, Y. and Kirch, K. and Knecht, A. and Knowles, P. and Koch, H.-C. and Koss, P. A. and Komposch, S. and Kozela, A. and Kraft, A. and Krempel, J. and Ku\ifmmode \acute{z}\else \'{z}\fi{}niak, M. and Lauss, B. and Lefort, T. and Lemi\`ere, Y. and Leredde, A. and Mohanmurthy, P. and Mtchedlishvili, A. and Musgrave, M. and Naviliat-Cuncic, O. and Pais, D. and Piegsa, F. M. and Pierre, E. and Pignol, G. and Plonka-Spehr, C. and Prashanth, P. N. and Qu\'em\'ener, G. and Rawlik, M. and Rebreyend, D. and Rien\"acker, I. and Ries, D. and Roccia, S. and Rogel, G. and Rozpedzik, D. and Schnabel, A. and Schmidt-Wellenburg, P. and Severijns, N. and Shiers, D. and Tavakoli Dinani, R. and Thorne, J. A. and Virot, R. and Voigt, J. and Weis, A. and Wursten, E. and Wyszynski, G. and Zejma, J. and Zenner, J. and Zsigmond, G.},
  journal = {Phys. Rev. Lett.},
  volume = {124},
  issue = {8},
  pages = {081803},
  numpages = {7},
  year = {2020},
  month = {Feb},
  publisher = {American Physical Society},
  doi = {10.1103/PhysRevLett.124.081803},
  url = {https://link.aps.org/doi/10.1103/PhysRevLett.124.081803}
}

@article{PhysRevLett.97.131801,
  title = {{Improved Experimental Limit on the Electric Dipole Moment of the Neutron}},
  author = {Baker, C. A. and Doyle, D. D. and Geltenbort, P. and Green, K. and van der Grinten, M. G. D. and Harris, P. G. and Iaydjiev, P. and Ivanov, S. N. and May, D. J. R. and Pendlebury, J. M. and Richardson, J. D. and Shiers, D. and Smith, K. F.},
  journal = {Phys. Rev. Lett.},
  volume = {97},
  issue = {13},
  pages = {131801},
  numpages = {4},
  year = {2006},
  month = {Sep},
  publisher = {American Physical Society},
  doi = {10.1103/PhysRevLett.97.131801},
  url = {https://link.aps.org/doi/10.1103/PhysRevLett.97.131801}
}

@article{electron_edm1,
	title = {Improved limit on the electric dipole moment of the electron},
	volume = {562},
	rights = {2018 Springer Nature Limited},
	issn = {1476-4687},
	url = {https://www.nature.com/articles/s41586-018-0599-8},
	doi = {10.1038/s41586-018-0599-8},
	pages = {355--360},
	number = {7727},
	journal = {Nature},
	author = {Andreev, V. and Ang, D. G. and DeMille, D. and Doyle, J. M. and Gabrielse, G. and Haefner, J. and Hutzler, N. R. and Lasner, Z. and Meisenhelder, C. and O’Leary, B. R. and Panda, C. D. and West, A. D. and West, E. P. and Wu, X. and {ACME Collaboration}},
	urldate = {2020-10-06},
	year = {2018-10},
	langid = {english},
	note = {Number: 7727
Publisher: Nature Publishing Group}
}

@article{electron_edm2,
	title = {{Order of Magnitude Smaller Limit on the Electric Dipole Moment of the Electron}},
	volume = {343},
	issn = {0036-8075},
	url = {https://science.sciencemag.org/content/343/6168/269},
	doi = {10.1126/science.1248213},
	pages = {269--272},
	number = {6168},
	journal = {Science},
	author = {Baron, J. and Campbell, W. C. and {DeMille}, D. and Doyle, J. M. and Gabrielse, G. and Gurevich, Y. V. and Hess, P. W. and Hutzler, N. R. and Kirilov, E. and Kozyryev, I. and O’Leary, B. R. and Panda, C. D. and Parsons, M. F. and Petrik, E. S. and Spaun, B. and Vutha, A. C. and West, A. D.},
	year = {2014}
}

@article{hg_edm,
  title = {{Reduced Limit on the Permanent Electric Dipole Moment of $^{199}\mathrm{Hg}$}},
  author = {Graner, B. and Chen, Y. and Lindahl, E. G. and Heckel, B. R.},
  journal = {Phys. Rev. Lett.},
  volume = {116},
  issue = {16},
  pages = {161601},
  numpages = {5},
  year = {2016},
  month = {Apr},
  publisher = {American Physical Society},
  doi = {10.1103/PhysRevLett.116.161601},
  url = {https://link.aps.org/doi/10.1103/PhysRevLett.116.161601}
}

@article{PhysRevLett.119.119901,
  title = {{Erratum: Reduced Limit on the Permanent Electric Dipole Moment of $^{199}\mathrm{Hg}$ [Phys. Rev. Lett. 116, 161601 (2016)]}},
  author = {Graner, B. and Chen, Y. and Lindahl, E. G. and Heckel, B. R.},
  journal = {Phys. Rev. Lett.},
  volume = {119},
  issue = {11},
  pages = {119901},
  numpages = {1},
  year = {2017},
  month = {Sep},
  publisher = {American Physical Society},
  doi = {10.1103/PhysRevLett.119.119901},
  url = {https://link.aps.org/doi/10.1103/PhysRevLett.119.119901}
}

@article{PRDThomas,
  title = {{Local Lorentz transformations and Thomas effect in general relativity}},
  author = {Silenko, Alexander J.},
  journal = {Phys. Rev. D},
  volume = {93},
  issue = {12},
  pages = {124050},
  numpages = {17},
  year = {2016},
  month = {Jun},
  publisher = {American Physical Society},
  doi = {10.1103/PhysRevD.93.124050},
  url = {https://link.aps.org/doi/10.1103/PhysRevD.93.124050}
}

@article{PRD2016,
  title = {{Manifestations of the rotation and gravity of the Earth in high-energy physics experiments}},
  author = {Obukhov, Yuri N. and Silenko, Alexander J. and Teryaev, Oleg V.},
  journal = {Phys. Rev. D},
  volume = {94},
  issue = {4},
  pages = {044019},
  numpages = {14},
  year = {2016},
  month = {Aug},
  publisher = {American Physical Society},
  doi = {10.1103/PhysRevD.94.044019},
  url = {https://link.aps.org/doi/10.1103/PhysRevD.94.044019}
}

@article{PRD2017,
  title = {General treatment of quantum and classical spinning particles in external fields},
  author = {Obukhov, Yuri N. and Silenko, Alexander J. and Teryaev, Oleg V.},
  journal = {Phys. Rev. D},
  volume = {96},
  issue = {10},
  pages = {105005},
  numpages = {18},
  year = {2017},
  month = {Nov},
  publisher = {American Physical Society},
  doi = {10.1103/PhysRevD.96.105005},
  url = {https://link.aps.org/doi/10.1103/PhysRevD.96.105005}
}

@article{PRD2007,
  title = {Equivalence principle and experimental tests of gravitational spin effects},
  author = {Silenko, Alexander J. and Teryaev, Oleg V.},
  journal = {Phys. Rev. D},
  volume = {76},
  issue = {6},
  pages = {061101},
  numpages = {5},
  year = {2007},
  month = {Sep},
  publisher = {American Physical Society},
  doi = {10.1103/PhysRevD.76.061101},
  url = {https://link.aps.org/doi/10.1103/PhysRevD.76.061101}
}
\end{document}